\documentclass[times,twocolumn]{aastex62}

\usepackage{CJK}
\usepackage{amsmath}
\usepackage{multirow}
\usepackage{cases}
\usepackage{graphicx}
\usepackage{subfigure}
\usepackage{natbib}
\usepackage{color}
\usepackage{tabularx}
\usepackage{bm}
\usepackage{threeparttable}
\usepackage{gensymb}

\newcommand{\HL}[1]{#1}
\newcommand{\thetae}{\theta_{\rm E}}

\newcommand{\pie}{\pi_{\rm E}}

\newcommand{\te}{t_{\rm E}}
\newcommand{\event}{OGLE-2016-BLG-0523}

\DeclareGraphicsExtensions{.pdf,.png,.jpg}

\shorttitle{}
\shortauthors{Yang et al.}

\begin{document}
\begin{CJK*}{UTF8}{gbsn}

\title{{\large OGLE-2016-BLG-0523: A Low-mass Brown Dwarf or super-Jupiter in a $>100$ au Orbit}}

\correspondingauthor{Hongjing Yang, Weicheng Zang}
\email{hongjing.yang@qq.com, zangweicheng@westlake.edu.cn}

\author[0000-0003-0626-8465]{Hongjing Yang (杨弘靖)}
\affiliation{Department of Astronomy, Westlake University, Hangzhou 310030, Zhejiang Province, China}

\author[0000-0001-6000-3463]{Weicheng Zang (臧伟呈)}
\affiliation{Department of Astronomy, Westlake University, Hangzhou 310030, Zhejiang Province, China}

\author[0000-0001-9823-2907]{Yoon-Hyun Ryu} 
\affiliation{Korea Astronomy and Space Science Institute, Daejeon 34055, Republic of Korea}

\author[0000-0001-5207-5619]{Andrzej Udalski}
\affiliation{Astronomical Observatory, University of Warsaw, Al. Ujazdowskie 4, 00-478 Warszawa, Poland}

\author[0000-0001-5651-9440]{Yuchen Tang (唐雨辰)}
\affiliation{Department of Astronomy, Westlake University, Hangzhou 310030, Zhejiang Province, China}

\collaboration{(Leading Authors)}

\author[0000-0003-3316-4012]{Michael D. Albrow}
\affiliation{University of Canterbury, School of Physical and Chemical Sciences, Private Bag 4800, Christchurch 8020, New Zealand}

\author[0000-0001-6285-4528]{Sun-Ju Chung}
\affiliation{Korea Astronomy and Space Science Institute, Daejeon 34055, Republic of Korea}

\author{Andrew Gould} 
\affiliation{Max-Planck-Institute for Astronomy, K\"onigstuhl 17, 69117 Heidelberg, Germany}
\affiliation{Department of Astronomy, Ohio State University, 140 W. 18th Ave., Columbus, OH 43210, USA}

\author[0000-0002-2641-9964]{Cheongho Han}
\affiliation{Department of Physics, Chungbuk National University, Cheongju 28644, Republic of Korea}

\author[0000-0002-9241-4117]{Kyu-Ha Hwang}
\affiliation{Korea Astronomy and Space Science Institute, Daejeon 34055, Republic of Korea}

\author[0000-0002-0314-6000]{Youn Kil Jung}
\affiliation{Korea Astronomy and Space Science Institute, Daejeon 34055, Republic of Korea}
\affiliation{National University of Science and Technology (UST), Daejeon 34113, Republic of Korea}

\author[0000-0002-4355-9838]{In-Gu Shin}
\affiliation{Department of Astronomy, Westlake University, Hangzhou 310030, Zhejiang Province, China}


\author[0000-0003-1525-5041]{Yossi Shvartzvald}
\affiliation{Department of Particle Physics and Astrophysics, Weizmann Institute of Science, Rehovot 7610001, Israel}

\author[0000-0001-9481-7123]{Jennifer C. Yee}
\affiliation{Center for Astrophysics $|$ Harvard \& Smithsonian, 60 Garden St., Cambridge, MA 02138, USA}

\author[0000-0002-7511-2950]{Sang-Mok Cha} 
\affiliation{Korea Astronomy and Space Science Institute, Daejeon 34055, Republic of Korea}
\affiliation{School of Space Research, Kyung Hee University, Yongin, Kyeonggi 17104, Republic of Korea} 

\author[0000-0002-4292-9649]{Dong-Jin Kim}
\affiliation{Korea Astronomy and Space Science Institute, Daejeon 34055, Republic of Korea}

\author[0000-0001-8263-1006]{Hyoun-Woo Kim}
\affiliation{Korea Astronomy and Space Science Institute, Daejeon 34055, Republic of Korea}

\author[0000-0003-0562-5643]{Seung-Lee Kim} 
\affiliation{Korea Astronomy and Space Science Institute, Daejeon 34055, Republic of Korea}

\author[0000-0003-0043-3925]{Chung-Uk Lee}
\affiliation{Korea Astronomy and Space Science Institute, Daejeon 34055, Republic of Korea}

\author[0009-0000-5737-0908]{Dong-Joo Lee} 
\affiliation{Korea Astronomy and Space Science Institute, Daejeon 34055, Republic of Korea}

\author[0000-0001-7594-8072]{Yongseok Lee} 
\affiliation{Korea Astronomy and Space Science Institute, Daejeon 34055, Republic of Korea}
\affiliation{School of Space Research, Kyung Hee University, Yongin, Kyeonggi 17104, Republic of Korea}

\author[0000-0002-6982-7722]{Byeong-Gon Park}
\affiliation{Korea Astronomy and Space Science Institute, Daejeon 34055, Republic of Korea}

\author[0000-0003-1435-3053]{Richard W. Pogge} 
\affiliation{Department of Astronomy, Ohio State University, 140 West 18th Ave., Columbus, OH  43210, USA}
\affiliation{Center for Cosmology and AstroParticle Physics, Ohio State University, 191 West Woodruff Ave., Columbus, OH 43210, USA}

\collaboration{(The KMTNet Collaboration)}

\author[0000-0001-7016-1692]{Przemek Mr\'{o}z}
\affiliation{Astronomical Observatory, University of Warsaw, Al. Ujazdowskie 4, 00-478 Warszawa, Poland}

\author[0000-0002-0548-8995]{Micha{\l}~K. Szyma\'{n}ski}
\affiliation{Astronomical Observatory, University of Warsaw, Al. Ujazdowskie 4, 00-478 Warszawa, Poland}

\author[0000-0002-2335-1730]{Jan Skowron}
\affiliation{Astronomical Observatory, University of Warsaw, Al. Ujazdowskie 4, 00-478 Warszawa, Poland}

\author[0000-0002-9245-6368]{Radoslaw Poleski}
\affiliation{Astronomical Observatory, University of Warsaw, Al. Ujazdowskie 4, 00-478 Warszawa, Poland}

\author[0000-0002-7777-0842]{Igor Soszy\'{n}ski}
\affiliation{Astronomical Observatory, University of Warsaw, Al. Ujazdowskie 4, 00-478 Warszawa, Poland}

\author[0000-0002-2339-5899]{Pawe{\l} Pietrukowicz}
\affiliation{Astronomical Observatory, University of Warsaw, Al. Ujazdowskie 4, 00-478 Warszawa, Poland}

\author[0000-0003-4084-880X]{Szymon Koz{\l}owski}
\affiliation{Astronomical Observatory, University of Warsaw, Al. Ujazdowskie 4, 00-478 Warszawa, Poland}

\author[0000-0002-9326-9329]{Krzysztof A. Rybicki}
\affiliation{Astronomical Observatory, University of Warsaw, Al. Ujazdowskie 4, 00-478 Warszawa, Poland}
\affiliation{Department of Particle Physics and Astrophysics, Weizmann Institute of Science, Rehovot 76100, Israel}

\author[0000-0002-6212-7221]{Patryk Iwanek}
\affiliation{Astronomical Observatory, University of Warsaw, Al. Ujazdowskie 4, 00-478 Warszawa, Poland}

\author[0000-0001-6364-408X]{Krzysztof Ulaczyk}
\affiliation{Department of Physics, University of Warwick, Gibbet Hill Road, Coventry, CV4~7AL,~UK}

\author[0000-0002-3051-274X]{Marcin Wrona}
\affiliation{Astronomical Observatory, University of Warsaw, Al. Ujazdowskie 4, 00-478 Warszawa, Poland}
\affiliation{Villanova University, Department of Astrophysics and Planetary Sciences, 800 Lancaster Ave., Villanova, PA 19085, USA}

\author[0000-0002-1650-1518]{Mariusz Gromadzki}
\affiliation{Astronomical Observatory, University of Warsaw, Al. Ujazdowskie 4, 00-478 Warszawa, Poland}

\author[0000-0002-8911-6581]{Mateusz J. Mr\'{o}z}
\affiliation{Astronomical Observatory, University of Warsaw, Al. Ujazdowskie 4, 00-478 Warszawa, Poland}

\collaboration{(The OGLE Collaboration)}

\author[0000-0002-1279-0666]{Jiyuan Zhang (张纪元)}
\affiliation{Department of Astronomy, Tsinghua University, Beijing 100084, China}

\author[0000-0001-8317-2788]{Shude Mao (毛淑德)}
\affiliation{Department of Astronomy, Westlake University, Hangzhou 310030, Zhejiang Province, China}

\author[0009-0001-6584-7187]{Hongyu Li (李弘禹)}
\affiliation{Department of Astronomy, Tsinghua University, Beijing 100084, China}

\author[0000-0003-4625-8595]{Qiyue Qian}
\affiliation{Department of Astronomy, Tsinghua University, Beijing 100084, China}

\begin{abstract}

To search for very wide-orbit binary-lens events, we cross-match microlensing events detected by OGLE-IV and KMTNet between 2011 and 2024. After removing contaminants from variable stars, we identify OGLE-2016-BLG-0523 as a microlensing event showing two short-lived bumps in its light curve, with the two separated by 8.36 yr. We investigate three scenarios to explain the two bumps, including a binary-lens single-source (2L1S), a single-lens binary-source (1L2S), and two unrelated single-lens single-source events (2$\times$1L1S). The 1L2S interpretation is excluded due to its poorer fit to the data, while the 2$\times$1L1S scenario is disfavored by the low event rate, unphysical total source flux, and unphysical color for one of the two putative sources. For the surviving 2L1S interpretation, the Einstein timescales of the two bumps are $8.343 \pm 0.076$ days and $3.231 \pm 0.047$ days, respectively, implying a binary mass ratio of $q = 0.15$. A Bayesian analysis indicates that the primary lens is most likely a low-mass M dwarf, while the companion is probably a low-mass brown dwarf or super-Jupiter. The inferred projected separation is about 450 au, making the companion the widest low-mass object discovered via microlensing to date. Our discovery highlights both the promise and challenges of identifying very wide, low-mass companions through microlensing. Although upcoming facilities such as Roman and Earth 2.0 space telescopes will improve sensitivity to such events, overcoming degeneracies and distinguishing them from variable phenomena will remain essential for building a robust statistical sample.

\end{abstract}

\section{Introduction}\label{intro}
The gravitational microlensing technique can detect low-mass objects, including brown dwarfs (BDs) and planets, around a wide range of host stars \citep{Shude1991,Andy1992}. It is most sensitive to objects located near the Einstein radius, with $s \sim 1$, where $s$ is the projected lens separation scaled to the Einstein radius. In this regime, the caustic structure is largest and the cross-section for a source to cross a caustic or approach a cusp is maximized \citep{Andy1992,OB190960}. For a typical host of half a solar mass at a distance of 5 kpc, the Einstein radius is about 2.8 au and the corresponding mean three-dimensional separation is about 3.1 au, implying an orbital period of roughly 9 years. Thus, microlensing is most sensitive to companions in Jupiter-like orbits.

For companions far beyond the Einstein radius ($s \gg 1$), the geometric detection probability decreases approximately as $1/s$. Among the more than 280 known microlensing planets \citep{NASAExo} \HL{as of 2026 September 25\footnote{\url{https://exoplanetarchive.ipac.caltech.edu/}}}, only $\sim 10$ cases unambiguously have $s > 2$ (e.g., \citealt{OB08092,OB160263,OB160007}). The study of companions at very wide orbits has two important implications. First, it enables us to investigate planets and brown dwarfs beyond Jupiter-like orbits, including icy worlds on Neptune-like orbits and even objects residing in ``exo-Kuiper'' and ``exo-Oort'' analogs \citep{GouldOort}. Second, microlensing can detect free-floating planets (FFPs), which are gravitationally unattached to any star \citep{Sumi2011,Mroz2017a,Gould2022_FFP_EinsteinDesert,Sumi2023}. However, events attributed to FFPs may instead be caused by planets that remain bound to distant hosts, for which the projected planet--host separation is so large that the host leaves no detectable signature in the microlensing light curve. Therefore, studying microlensing planets at very wide orbits and comparing their occurrence rate \citep{OGLE_wide} with that of FFPs may constrain the fraction of FFP events that are actually due to bound planets.

The widest separation of a known low-mass companion detected by microlensing is $s=5.26\pm0.11$, with a binary mass ratio of $q=(2.41\pm0.45)\times10^{-4}$ \citep{OB08092}. Some other events admit solutions with larger $s$, but the well-known close/wide degeneracy, in which the lens geometry is approximately related by the transformation $s\leftrightarrow s^{-1}$ \citep{Griest1998}, implies that an $s\ll1$ solution cannot be excluded (e.g., \citealt{KB200414}).

Searching for well-separated bumps (i.e., repeating events) provides a channel to detect low-mass companions at very wide orbits \citep{Di1996_repeatevent,Hanrepeating,OB160263}. Recently, \cite{OGLE_wide_binary} reported the discovery of 107 binary events with well-separated bumps from the 2010--2019 data of the fourth phase of the Optical Gravitational Lensing Experiment (OGLE-IV, \citealt{OGLEIV}), \HL{which monitored the event with the 1.3\,m Warsaw Telescope at Las Campanas Observatory in Chile.} Of these, 59 are most likely binary lenses and 48 are best interpreted as binary sources. Among the 59 probable binary-lens events, none of the events with $s>10$ have a short-duration bump that is characterized by an event timescale $\te<5\,$days.  

To search for very wide-orbit binary-lens events through repeating microlensing events, we cross-matched all events discovered by the OGLE-IV \HL{Early Warning System}\footnote{\url{https://ogle.astrouw.edu.pl/ogle4/ews/ews.html}} \citep{Udalski1994, Udalski2003} from 2011 to 2024 and by the Korea Microlensing Telescope Network (KMTNet, \citealt{KMT2016})\footnote{\url{https://kmtnet.kasi.re.kr/ulens/}} \HL{EventFinder system} \citep{KMTeventfinder} from 2016 to 2024. \HL{KMTNet observations were obtained with its network of three 1.6\,m telescopes located at Cerro Tololo Inter-American Observatory in Chile (KMTC), the South African Astronomical Observatory in South Africa (KMTS), and Siding Spring Observatory in Australia (KMTA).} We found 31 pairs of events whose cataloged positions are within $2\arcsec$ of each other. Among these, three pairs have at least one event for which the OGLE or KMTNet automatic point-source point-lens (PSPL) fit yields an event timescale of $\te < 5$ days. These are OGLE-2016-BLG-0523/KMT-2016-BLG-0461/KMT-2024-BLG-2084, KMT-2016-BLG-0662/OGLE-2024-BLG-0548, and KMT-2016-BLG-1704/KMT-2024-BLG-1787. We then performed tender-loving-care (TLC) re-reduced photometry and found that the $I$- and $V$-band light curves of the latter two pairs show features characteristic of cataclysmic variables 
\HL{(their light curves are shown in Appendix \ref{rej-lc}).}
The first one, OGLE-2016-BLG-0523/KMT-2016-BLG-0461/KMT-2024-BLG-2084, is confirmed to be a microlensing event. The two microlensing signatures in the light curve are separated by 8.36~yr, and a binary-lens single-source (2L1S) model yields $s \sim 340$, roughly two orders of magnitude larger than the widest case previously reported by \cite{OB08092}.

In this work we carry out a detailed study of OGLE-2016-BLG-0523/KMT-2016-BLG-0461/KMT-2024-BLG-2084. Section~\ref{obser} introduces the OGLE and KMTNet observations and data reductions for this event. In Sections~\ref{model} and \ref{2-1L1S}, we analyze the data with different lensing models. Section~\ref{lens} estimates the physical properties of the lens system. Finally, Section~\ref{dis} discusses the implications of this discovery.

\section{Observations}\label{obser}
\begin{figure*}
    \centering
    \includegraphics[width=\linewidth]{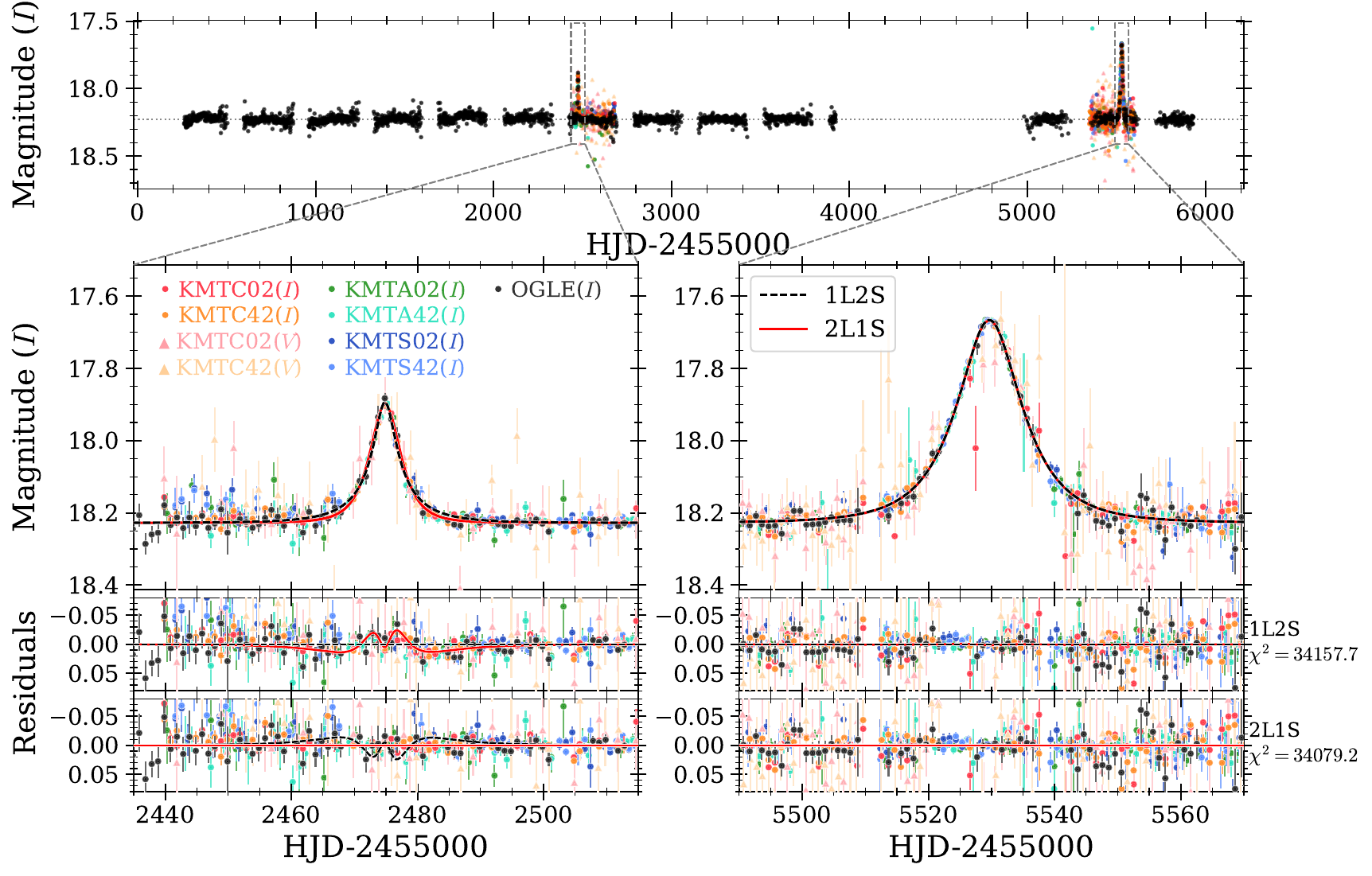}
    \caption{Light curve of the microlensing event \event, showing the 1L2S model (dashed magenta line) and the 2L1S model (solid black line). Different datasets are shown with distinct colors and markers, and data are binned daily for clarity. The upper panel displays the full light curve from 2010 to 2025, while the lower panels zoom in on the two events in 2016 and 2024 with their residuals. In each residual panel, we show both the residuals relative to that model and the difference between the 2L1S and 1L2S models for comparison. The baseline magnitude, $I = 18.226$~\HL{mag}, is derived from the CFHT images with PSF of $\sim 0.''50$.}
    \label{fig:lc}
\end{figure*}

The event was initially identified by the OGLE Early Warning System in 2016 and was designated as \event. The event is located at equatorial coordinates $(\alpha, \delta)_{\rm J2000}$ = (17:56:52.92, $-$28:34:12.4) and Galactic coordinates $(\ell, b) = (1.5763, -1.9346)$. Independently, the event was recovered in the KMTNet post-season search using the EventFinder algorithm and received the designation as KMT-2016-BLG-0461. The second microlensing brightening in 2024 was later reported by the KMTNet AlertFinder system \citep{KMTAF} on 2024 August 05 and was labeled as KMT-2024-BLG-2084. Throughout the remainder of this work, we adopt the earliest reported designation, \event, to refer to the event.

\event\ lies in the OGLE BLG504 field. This field was monitored at a cadence of approximately $4\ \mathrm{hr}^{-1}$ during the 2016 season and at about $1\ \mathrm{hr}^{-1}$ in 2024. The position of \event\ also places it within the overlapping region of the KMTNet BLG02 and BLG42 fields, which together provide a cadence of $4\ \mathrm{hr}^{-1}$ in 2016 and of $4\ \mathrm{hr}^{-1}$ for KMTC and $6\ \mathrm{hr}^{-1}$ for KMTS and KMTA in 2024. The layout of the KMTNet fields can be found in Figure~12 of \citet{KMTeventfinder}. For both OGLE and KMTNet, the majority of the images were obtained in the $I$ band, while a fraction of exposures were secured in the $V$ band to determine the source color.

All photometry used in the modeling was obtained through difference imaging analysis (DIA; \citealt{Tomaney1996,Alard1998}), using the implementations developed independently by the OGLE and KMTNet teams. The OGLE pipeline is described in \citet{Wozniak2000}, and the KMTNet reductions are based on \citet{pysis} together with the TLC reduction procedures of \citet{Yang_TLC, Yang_TLC2}. The OGLE data consist of a single $I$-band dataset, OGLE ($I$), spanning the interval from 2010 to 2025. For KMTNet, the distortion patterns differ substantially between the 2016 and 2024 seasons. We therefore process images from these two years separately for each of the overlapping BLG02 and BLG42 fields. 

\HL{For each data set, the photometric error bars of individual points estimated by the DIA pipelines, $\sigma_0$, were re-scaled following the prescription described in \cite{MB11293}, $\sigma = k\sqrt{\sigma_{\rm min}^2 + \sigma_0^2}$, where $\sigma_{\rm min}$ accounts for the minimum photometric uncertainty and $k$ is a rescaling factor. The values of $\sigma_{\rm min}$ and $k$ were adjusted so that each data set satisfies $\chi^{2}/{\rm dof}=1$.}

In addition to the $I$-band data, we include the KMTC02 and KMTC42 $V$-band light curves to determine the source colors for the single-lens binary-source (1L2S; \citealt{Gaudi1998}) model. These $V$-band datasets are denoted KMTC02-16 ($V$), KMTC02-24 ($V$), KMTC42-16 ($V$), and KMTC42-24 ($V$). The 1L2S fit requires the flux ratio between the two sources, which in turn requires that the source fluxes from the 2016 and 2024 data sets be placed on a common magnitude system. For the $I$-band, the OGLE photometry was from a single reduction, so the $I$-band flux ratio is obtained directly from the OGLE source fluxes. For the $V$-band, however, the 2016 and 2024 KMTNet reductions were performed independently. We therefore matched common field stars within $1'$ of the event between the KMTC42-16 ($V$) and KMTC42-24 ($V$) datasets and found
\begin{equation}
    V_{\rm KMTC42,24} - V_{\rm KMTC42,16} = 0.24 \pm 0.01~{\rm mag},
\end{equation}
where $V_{\rm KMTC42,24}$ and $V_{\rm KMTC42,16}$ are the instrumental $V$ magnitudes in the KMTC42-24 and KMTC42-16 reductions, respectively. We impose the following prior on the relative calibration of the two KMTC42 source fluxes:
\begin{equation}
    \mathcal{L}_{\rm prior}(V) = \exp\left(-\frac{(0.24 + 2.5 \times \log{(\frac{f_{\rm S, KMTC42,24}}{f_{\rm S, KMTC42,16}}}))^2}{0.01^2}\right),
\end{equation}
where $f_{\rm S,KMTC42,24}$ and $f_{\rm S,KMTC42,16}$ denote the total source fluxes (instrumental units) in the KMTC42-24 and KMTC42-16 datasets, respectively.

\HL{The name of each data set, filter, date coverage, number of data points, and error renormalization factors are summarized in Table~\ref{tab:data}.}

\section{Standard Modeling: 2L1S and 1L2S}\label{model}
If the two bumps observed in 2016 and 2024 arise from a single microlensing event, the photometric behavior of \event\ can be explained either by a binary-lens single-source configuration (2L1S), in which the two PSPL-like features originate from a common source star, or by a single-lens two-source configuration (1L2S), in which a single lens magnifies two different source stars. Thus, in this section we fit the data using both the 2L1S and 1L2S models.

 \begin{table*}[htb]
    \renewcommand\arraystretch{1.20}
    \centering
    \caption{Lensing Parameters for 2L1S, 1L2S and 2$\times$1L1S Models}
    \begin{tabular}{c c|c c|c c}
    \hline
    \hline
    \multicolumn{2}{c|}{2L1S} & \multicolumn{2}{c|}{1L2S} & \multicolumn{2}{c}{2$\times$1L1S}  \\
    Parameters & Value & Parameters & Value & Parameters & Value \\
    \hline
    $\chi^2$/dof & 34079.2/34080 & $\chi^2$/dof & 34157.7/34079 & $\chi^2$/dof & 34118.6/34078  \\
    \hline
    $t_{0,\mathrm{L1}}-2450000$ & $10529.652 \pm 0.021$ & $t_{0,\mathrm{S1}}-2450000$ & $10529.653 \pm 0.021$ & $t_{0,\mathrm{L1S1}}-2450000$ & $10529.655 \pm 0.022$ \\
    
    $u_{0,\mathrm{L1}}$ & $0.5991 \pm 0.0075$ & $u_{0,\mathrm{S1}}$ & $0.5624 \pm 0.0049$ & $u_{0,\mathrm{L1S1}}$ & $0.5226 \pm 0.0104$ \\
    
    $t_{\mathrm{E,L1}}$ (day) & $8.343 \pm 0.076$ & $t_{0,\mathrm{S2}}-2450000$ & $7474.850 \pm 0.033$ & $t_{\mathrm{E,L1S1}}$ (day) & $9.103 \pm 0.130$ \\
    
    $t_{0,\mathrm{L2}}-2450000$ & $7474.876 \pm 0.030$ & $u_{0,\mathrm{S2}}$ & $0.1903 \pm 0.0057$ & $t_{0,\mathrm{L2S2}}-2450000$ & $7474.874 \pm 0.036$ \\ 
    
    $u_{0,\mathrm{L2}}$ & $0.8583 \pm 0.0100$ & $t_{\mathrm{E}}$ (day) & $8.668 \pm 0.058$ & $u_{0,\mathrm{L2S2}}$ & $0.3023 \pm 0.0348$ \\
    
    $t_{\mathrm{E,L2}}$ (day) & $3.231 \pm 0.047$ & $q_{f,I}$ &  $0.1219 \pm 0.0049$ & $t_{\mathrm{E,L2S2}}$ (day) & $6.043 \pm 0.349$ \\  
    
    $q$ & $0.1497 \pm 0.0042$ & $q_{f,V}$ &  $0.1270 \pm 0.0027$ & $q_{f,I}$ & $0.2617 \pm 0.0149$ \\
    
    $s$ & $341.0 \pm 3.0$ & &  & $q_{f,V}$ &  $0.2680 \pm 0.0143$ \\
    
    $I_{\rm S, OGLE}$ (\HL{mag}) & $18.223 \pm 0.022$ & $I_{\rm S, OGLE}$ (\HL{mag}) & $18.212 \pm 0.014$ & $I_{\rm S, OGLE}$ (\HL{mag}) & $18.210 \pm 0.013$ \\
    \hline
    \hline
    \end{tabular}
    \tablecomments{$I_{\rm S, OGLE}$ is the total source flux calibrated to the OGLE-III magnitude system. $q$ and $s$ of the 2L1S model are derived by Equation~(\ref{equ:qs}). \HL{The $t_0$ values with different subscripts are given in Heliocentric Julian Date.}}
    \label{tab:parm}
\end{table*}

\subsection{2L1S Modeling}\label{sec:2L1S}

For the 2L1S interpretation, the two PSPL features are sufficiently separated in time that, during each peak, the magnification can be treated as arising effectively from a single lens. Hence our 2L1S model is described by six parameters, consisting of two sets of PSPL parameters, $(t_{0,\mathrm{L1}},\,u_{0,\mathrm{L1}},\,t_{\mathrm{E,L1}})$ and $(t_{0,\mathrm{L2}},\,u_{0,\mathrm{L2}},\,t_{\mathrm{E,L2}})$. Here $t_0$ denotes the time of closest approach of the source to the lens, $u_0$ is the impact parameter in units of the angular Einstein radius $\thetae$, and $\te$ is the Einstein timescale. Subscripts L1 and L2 refer to the primary and secondary lenses, respectively; we assign L1 to the longer event (the 2024 peak) and L2 to the shorter event (the 2016 peak). Then, the binary mass ratio and the lens projected separation can be derived by 
\begin{equation}\label{equ:qs}
    q = \frac{t_{\mathrm{E,L2}}^2}{t_{\mathrm{E,L1}}^2}; \qquad
    s = \frac{t_{0,\mathrm{L1}} - t_{0,\mathrm{L2}}}{\sqrt{t_{\mathrm{E,L2}}^2 + t_{\mathrm{E,L1}}^2}}.
\end{equation}

The classical 2L1S model includes a seventh parameter, $\rho$, the ratio of the angular source radius to the angular Einstein radius, to account for finite-source effects \citep{1994ApJ...421L..75G,Shude1994,Nemiroff1994}. In the present case the light curves show no sign of finite-source features. Moreover, from Sections~\ref{lens} and ~\ref{dis} the angular source radius is $\sim1\ \mu$as while the typical angular Einstein radius is $\sim100\ \mu$as, implying $\rho\sim10^{-2}$. Given that the impact parameters for both lenses are of order $u_0\sim1$, finite-source effects are expected to be negligible. We therefore omit $\rho$ from the fits.

For each data set $i$, we include two flux parameters, $(f_{{\rm S},i},\,f_{{\rm B},i})$, which represent the source flux and the blended flux, respectively. For short-duration events the source and blend fluxes are often poorly constrained and can lead to an unphysically large negative blend. To avoid this, we check the $i'$-band baseline images taken by the 3.6~m Canada-France-Hawaii Telescope (CFHT), with PSF of $\sim 0.''50$. We calibrate the CFHT $i'$-band photometry to the OGLE-III $I$ system \citep{OGLEIII} using field stars within $1'$ of the event and find a star at the event position with $I=18.226\pm0.013$~\HL{mag}. We also calibrate the KMTC42-24 ($I$) dataset to the OGLE-III system and impose the following prior on its blend flux $f_{\rm B}$:
\begin{equation}\label{equ:blend}
\mathcal{L}_{\rm prior}(f_{\rm B})=
\begin{cases}
1, & f_{\rm B}\ge 0,\\[4pt]
\exp\left(-\dfrac{f_{\rm B}^2}{2\sigma^2}\right), & f_{\rm B}<0,
\end{cases}
\end{equation}
where $\sigma$ is the flux uncertainty corresponding to the calibrated magnitude $I=18.226\pm0.013$~\HL{mag}. 

We first fit the 2016 and 2024 datasets individually to obtain initial estimates of the two sets of PSPL parameters. For each season we sample the parameter posterior using Markov chain Monte Carlo (MCMC) with the \texttt{emcee} ensemble sampler \citep{emcee}, and use the resulting posterior medians as starting points for a global fit to all data. Final best-fit models were obtained by a downhill minimization\footnote{We use the Nelder-Mead simplex implementation in \textsc{SciPy}; see \url{https://docs.scipy.org/doc/scipy/reference/generated/scipy.optimize.fmin.html\#scipy.optimize.fmin}.}. Table~\ref{tab:parm} lists the resulting 2L1S parameters with their $1\sigma$ uncertainties derived from the MCMC posteriors.

The short timescale of the secondary lens, $t_{\mathrm{E,L2}} = 3.231 \pm 0.047\ \mathrm{day}$, indicates a low-mass object in the lens system. The separation inferred from Equation~(\ref{equ:qs}) is $s = 341.0 \pm 3.0$, implying a system that is significantly more widely separated than any previously reported cases with a low-mass object. We note that, even among known microlensing binary-lens systems, the widest previously reported system was from the OGLE event BLG654.29.7310 \citep{OGLE_wide_binary}, with $s \sim 35$ and $q \sim 0.68$.

\subsection{1L2S Modeling}\label{sec:1L2S}

For a 1L2S model, the total effective magnification in a given band $\lambda$ is \citep{MB12486}
\begin{equation}\label{equ:1L2S}
    A_{\lambda}(t)
    = \frac{A_{1}(t)\,f_{{\rm S},1,\lambda} + A_{2}(t)\,f_{{\rm S},2,\lambda}}
           {f_{{\rm S},1,\lambda} + f_{{\rm S},2,\lambda}}
    = \frac{A_{1}(t) + q_{f,\lambda}\,A_{2}(t)}{1 + q_{f,\lambda}},
\end{equation}
where
\begin{equation}
    q_{f,\lambda} \equiv \frac{f_{{\rm S},2,\lambda}}{f_{{\rm S},1,\lambda}} .
\end{equation}\label{eq:qF}
Here $f_{{\rm S},j,\lambda}$ denotes the source flux of source $j$ in band $\lambda$, and $A_{j}(t)$ is the PSPL magnification of source $j$ ($j=1,2$ for the primary and secondary sources, respectively). The parameter $q_{f,\lambda}$ is therefore the flux ratio of the secondary to the primary source in band $\lambda$.

Under the 1L2S interpretation the two microlensing signatures (2016 and 2024) share the same Einstein timescale, which we denote by $t_{\rm E}$. The lens-source geometries for the two sources are described by two pairs $(t_{0,\mathrm{S1}},u_{0,\mathrm{S1}})$ and $(t_{0,\mathrm{S2}},u_{0,\mathrm{S2}})$. We label S1 for the 2024 peak (primary) and S2 for the 2016 peak (secondary). Including the two flux-ratio parameters, $q_{f,I}$ and $q_{f,V}$, the 1L2S model therefore contains seven lensing parameters in total. 

The 1L2S lensing parameters derived from the MCMC analysis are listed in Table~\ref{tab:parm}. Compared to the 2L1S model, the 1L2S model is disfavored by $\Delta\chi^2 = 78.5$, which is sufficiently significant to reject the 1L2S interpretation. We find that the 1L2S model provides a worse fit to both bumps, with most of the $\Delta\chi^2$ ($\sim 60$) arising from the 2016 peak. As shown in Figure~\ref{fig:lc}, the most pronounced difference between the 2L1S and 1L2S models occurs in the wings of the 2016 peak, which contributes a large fraction of the total $\Delta\chi^2$. We therefore exclude the 1L2S scenario.

\section{Two Independent Microlensing Events?}\label{2-1L1S}
We have investigated two possible interpretations in which the two bumps observed in 2016 and 2024 are physically related: (1) they share the same source but are caused by different lenses (2L1S interpretation), and (2) they share the same lens but involve different source stars (1L2S interpretation). However, it is also possible that the two bumps are unrelated and thus correspond to two independent microlensing events (2$\times$1L1S interpretation).

\subsection{Source Position}

To test this interpretation, we first examine the consistency between the source positions inferred from the 2016 and 2024 bumps. For the OGLE data, we measure the source positions using difference imaging of four exposures at the peak of the 2016 bump and five exposures at the peak of the 2024 bump. This yields an offset between the two source positions of $\theta_{\rm S, 2024}-\theta_{\rm S, 2016} ({\rm N, E}) = (-52 \pm 35, 8 \pm 43)~{\rm mas}$.

However, this offset does not account for the proper motion of the source star or the reference field stars used for astrometric calibration between the 2016 and 2024 images. According to the OGLE-URANUS astrometry database (Udalski et al., in prep.), the source has a heliocentric proper motion of $\mu_{\rm S} ({\rm N, E}) = (-7.21 \pm 0.17, -4.91 \pm 0.24)~{\rm mas~yr^{-1}}$. As described in the Appendix, bulge red clump (RC) stars exhibit a mean proper motion of $\mu_{\rm RC} ({\rm N, E}) = (-5.34 \pm 0.12, -2.91 \pm 0.12)~{\rm mas~yr^{-1}}$ based on \textit{Gaia} DR3 \citep{GaiaDR3}. Because bulge stars dominate the astrometric reference frame, we adopt $\mu_{\rm RC}$ as the reference and correct the measured offset accordingly. The resulting corrected offset between the 2016 and 2024 source positions is
\begin{equation}
\theta_{\rm S, 2024}-\theta_{\rm S, 2016} ({\rm N, E}) = (-36 \pm 35, 24 \pm 43)~{\rm mas}~({\rm OGLE}).
\end{equation}

We also analyze the KMTC42 images, which provide the highest image quality and highest signal-to-noise ratio among the KMTNet data sets. The source positions are derived following the procedure described in \citet{Yang_TLC}. Applying the same proper-motion correction, we obtain an offset between the 2016 and 2024 source positions of
\begin{equation}
\theta_{\rm S, 2024}-\theta_{\rm S, 2016} ({\rm N, E}) = (4 \pm 69, 41 \pm 68)~{\rm mas}~({\rm KMTC42}).
\end{equation}
Therefore, both the OGLE and KMTNet measurements suggest that the source positions of the two bumps are consistent within the $\lesssim 1\sigma$ level in each coordinate direction. 

For a conservative estimate, we consider the probability that the two bumps are produced by two unrelated microlensing events separated by $\leq 100$ mas. The KMTNet prime fields cover $\sim 13~\mathrm{deg}^2$ and contain about 1120 events per year, corresponding to a microlensing event rate of $\sim 86~{\rm deg^{-2}~yr^{-1}}$ according to the KMTNet EventFinder system. Over our eight-year baseline, this corresponds to $\sim 8 \times 1120 \simeq 8960$ events in the prime fields. The probability of having two unrelated events within an angular separation of 100 mas is then roughly
\begin{equation}
\begin{split}
P_{\rm prime} &= 0.5\times(86~{\rm deg^{-2}~yr^{-1}})\times (8~{\rm yr}) \times (8960) \\
&\quad \times \pi (100~{\rm mas})^2 \sim 7.5 \times 10^{-3}.
\end{split}
\end{equation}
The KMTNet sub-prime fields cover $\sim 84~\mathrm{deg}^2$ and contain about 1980 events per year, yielding an event rate of $\sim 24~{\rm deg^{-2}~yr^{-1}}$. Applying the same calculation, we obtain $P_{\rm subprime} \sim 3.7 \times 10^{-3}$, and thus the combined probability is $P \sim 1.1\times10^{-2}$. If we further require that one of the events has a short timescale ($t_{\rm E} < 5$ days), the rate is reduced by a factor of $\sim 10$, based on the timescale distribution of \citet{Mroz2017a}, yielding $P \sim 1.1 \times 10^{-3}$. This implies a very low probability for two unrelated events. However, the corresponding statistical significance is only at the level of $\sim 2.6\sigma$ (or $\sim 3.0\sigma$ when including the short-timescale constraint), which is not sufficient to conclusively rule out this scenario. We therefore consider additional diagnostics.

\subsection{2$\times$1L1S Modeling}

Here, we test whether the two unrelated-event scenario is consistent with the light-curve modeling. Compared to the 2L1S model described in Section \ref{sec:2L1S}, the two lenses in the 2$\times$1L1S model have two different source stars, respectively, so the model requires two additional flux ratio parameters, $q_{f, I}$ and $q_{f, V}$, to characterize the two source and their definition is similar to Equation (\ref{eq:qF}). Compared to the 1L2S model described in Section \ref{sec:1L2S}, the two events are instead produced by different lenses, and thus the model requires one additional timescale parameter. 

We define the parameters of the 2$\times$1L1S model as follows, including two sets of PSPL parameters, ($t_{0,\mathrm{L1S1}}$, $u_{0,\mathrm{L1S1}}$, $t_{\mathrm{E,L1S1}}$) for the 2024 event and ($t_{0,\mathrm{L2S2}}$, $u_{0,\mathrm{L2S2}}$, $t_{\mathrm{E,L2S2}}$) for the 2016 event, as well as two flux-ratio parameters, $q_{f, I}$ and $q_{f, V}$. We also include the blended-flux constraint given in Equation (\ref{equ:blend}). The resulting parameters are listed in Table~\ref{tab:parm}. The 2$\times$1L1S model yields a worse fit than the 2L1S model by $\Delta\chi^2 = 39.4$.

Because the 2L1S model is a special case of the 2$\times$1L1S model, in which the two sources have identical fluxes, while the total source flux is twice that of the 2L1S model, one would not expect the 2$\times$1L1S model to provide a worse fit than the 2L1S model in the absence of the blended-flux constraint. Therefore, the worse fit is likely driven by the blended-flux constraint. Indeed, independent 1L1S fits to the 2016 and 2024 data yield a total source \HL{brightness} of $I_{\rm S} = 17.403 \pm 0.145$~\HL{mag}, which is approximately twice the total flux measured at the event position from CFHT imaging. This discrepancy strongly disfavors the 2$\times$1L1S model.

\subsection{Source Positions in a Color-Magnitude Diagram}

\begin{figure}[htb] 
    \centering
    \includegraphics[width=0.47\textwidth]{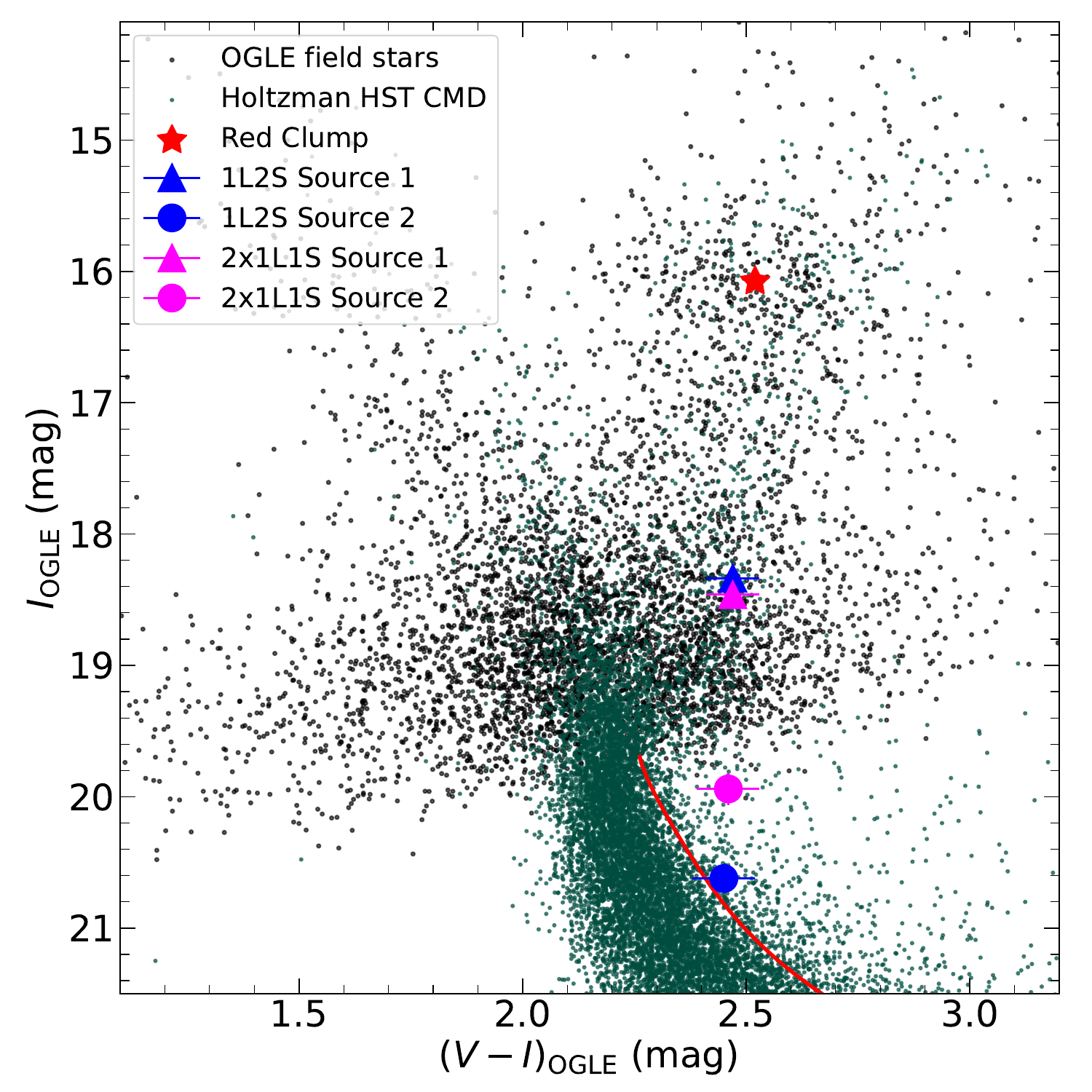}
    \caption{Color-magnitude diagram for \event. Black dots show OGLE-III field stars \citep{OGLEIII} within $1.5'$ of the event position. The red asterisk marks the red-clump centroid. The blue dot and triangle indicate the primary and secondary sources of the 1L2S model, respectively, while the magenta dot and triangle indicate the corresponding sources of the 2$\times$1L1S model. Green points show the \textit{HST} CMD from \cite{HSTCMD}, with the HST red-clump centroid shifted to match the OGLE-III red clump. The red line represents the red boundary of bulge main-sequence stars derived from stellar isochrones with [M/H] = $+0.5$ and an age of 5~Gyr. The secondary source is about $0.6\sigma$ redder than this boundary.}
    \label{fig:cmd}
\end{figure}

For many cases, the 1L2S scenario is tested by measuring the colors of the two sources and checking whether they are physically plausible \citep{Gaudi1998}. Here, we perform a similar test for the two sources in the 2$\times$1L1S model. From the 2$\times$1L1S fit, the source properties are
\((V-I,I)_{\rm S1}=(2.47\pm0.06,\;18.461\pm0.033)\)~\HL{mag} and
\((V-I,I)_{\rm S2}=(2.46\pm0.07,\;19.941\pm0.123)\)~\HL{mag}.
We place the two sources on a color-magnitude diagram constructed from OGLE-III field stars \citep{OGLEIII} within $1.5'$ of the event (see Figure~\ref{fig:cmd}; \citealt{Yoo2004}). Using the method of \citet{Nataf2013}, we measure the red-clump centroid to be
\((V-I,I)_{\rm cl}=(2.52\pm0.01,\;16.07\pm0.02)\)~\HL{mag}.
Adopting the intrinsic color and magnitude of the red clump, $(V-I,I)_{\rm cl,0}=(1.06\pm0.03,\;14.38\pm0.04)$~\HL{mag}, from \citet{Bensby2013} and Table~1 of \citet{Nataf2013}, we derive an extinction toward this line of sight of $A_I=1.69\pm0.04$~\HL{mag}.

We then register the \textit{HST} CMD of \citet{HSTCMD} to the OGLE-III frame using the \cite{MB07192} \textit{HST} red-clump centroid \((V-I,I)_{\rm cl,HST}=(1.62,\;15.15)\)~\HL{mag}. As illustrated in Fig.~\ref{fig:cmd}, the source for the first 1L1S event (S1) is consistent with a typical bulge subgiant, while the source for the second 1L1S event (S2) is redder than the bulge main-sequence population. To quantify the discrepancy, we estimate the red boundary of the bulge main sequence using PARSEC stellar isochrones \citep{PARSEC} with [M/H] = $+0.5$ and an age of 5~Gyr. For $3.5 \leq M_I \leq 10$~\HL{mag}, the red boundary of the selected isochrone is well fitted by \HL{the following polynomial function:}
\begin{equation}
(V-I) = \sum_{i=0}^{7} a_i (M_I-7)^i,
\end{equation}\label{equ:VIM}
where $M_I$ is the absolute magnitude in the $I$ band, \HL{and the coefficients of the polynomial $a_i$ are given in Table~\ref{tab:VIM}}. We show this relation as the red boundary in Fig.~\ref{fig:cmd}. We find that S2 is redder than the red boundary by $2.5\sigma$, and therefore physically implausible.

Combining the low event rate, the worse fit caused by the unphysical total source flux and the unphysical color for S2, we reject the interpretation that the two bumps are caused by two independent microlensing events and henceforth restrict consideration to the 2L1S model.

\section{Lens Properties from the Bayesian Analysis}\label{lens}
\begin{figure*}[htb] 
    \centering
    \includegraphics[width=0.85\textwidth]{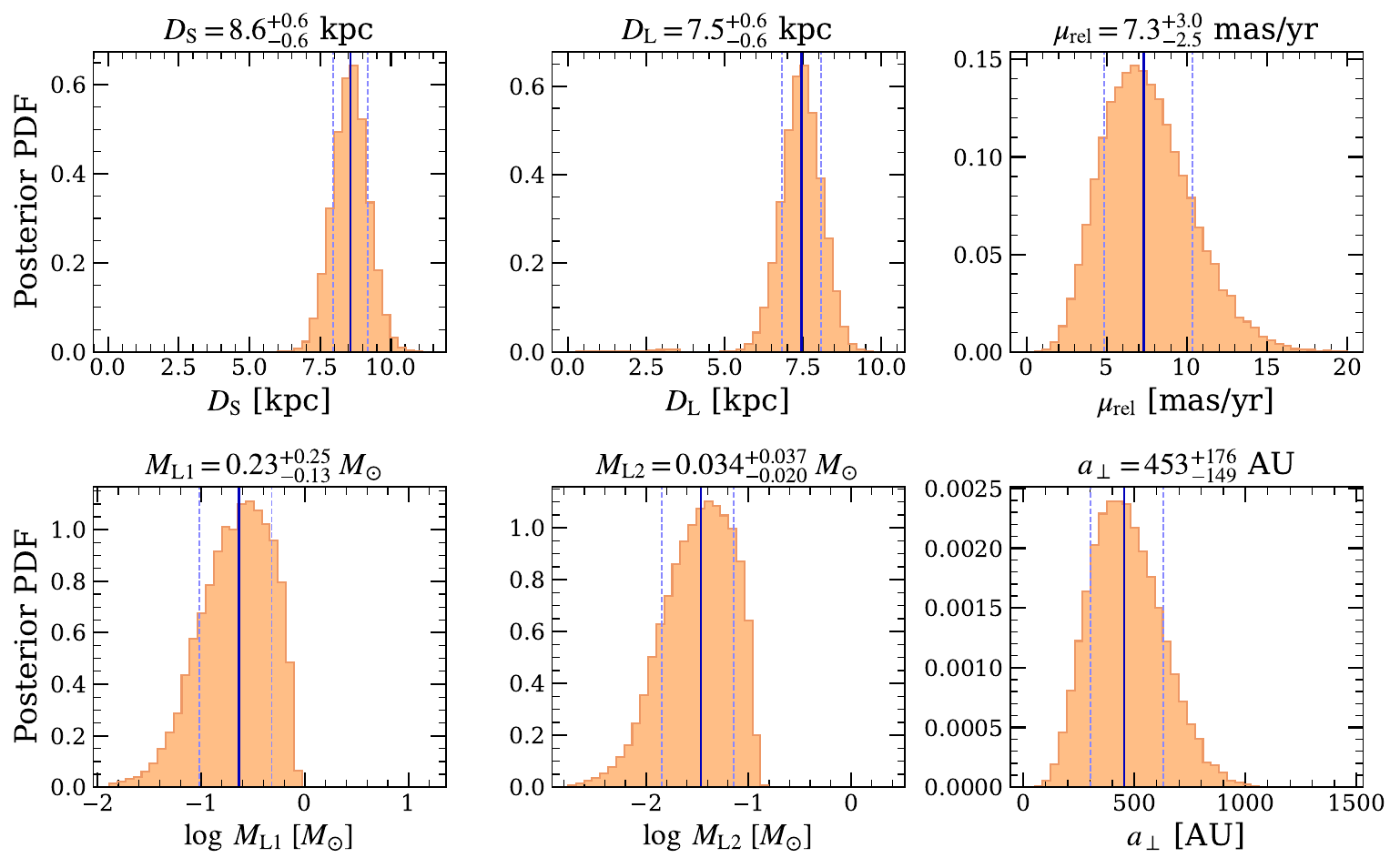}
    \caption{Posterior probability distributions of the lens and source physical parameters from the Bayesian analysis. The panels, from top to bottom and left to right, are the source distance $D_{\rm S}$, lens distance $D_{\rm L}$, lens-source relative proper motion $\mu_{\rm rel}$, the mass of the first and second lens $M_{\rm L1}$ and $M_{\rm L2}$, and their projected seperation $a_\perp$. The median and $\pm1\sigma$ values are marked in solid black and dashed blue vertical lines and are labeled on the top of each panel.}
    \label{fig:baye}
\end{figure*}

The lens mass $M_{\rm L}$ and distance $D_{\rm L}$, are linked to the angular Einstein radius $\thetae$ and the microlensing parallax $\pie$ as \citep{Gould1992,Gould2000}
\begin{equation}\label{eq:mass}
    M_{\rm L} = \frac{\theta_{\rm E}}{\kappa \pi_{\rm E}}, \qquad 
    D_{\rm L} = \frac{\mathrm{au}}{\pi_{\rm E}\theta_{\rm E} + \mathrm{au}/D_{\rm S}},
\end{equation}
where $D_{\rm S}$ denotes the distance to the source. Because our event yields no useful constraints on $\thetae$ or $\pie$, we infer the lens physical properties via a Bayesian framework that incorporates a Galactic prior.

Our Galactic prior is built from three ingredients: the lens mass function, the spatial density distributions of lenses and sources, and the kinematic distributions. For the mass function we adopt the \citet{Kroupa2001} initial mass function, applying upper-mass cutoffs of $1.3\,M_\odot$ for disk lenses and $1.1\,M_\odot$ for bulge lenses following \citet{Zhu2017spitzer}. The stellar density laws for both the disk and bulge are taken from \citet{Yang2021_GalacticModel}. Lens bulge velocities are modeled as in \citet{Zhu2017spitzer}, the source proper motion is adopted the measurement from the OGLE-URANUS system, and the disk kinematics follow the ``Model C'' of \citet{Yang2021_GalacticModel}, implemented using the \texttt{galpy} package \citep{bovy2015galpy}.

We generate a synthetic ensemble of $10^8$ microlensing trials. For each simulated event $i$, described by $(t_{{\rm E},i},\ \mu_{{\rm rel},i},\ \theta_{{\rm E},i})$, we compute a weight
\begin{equation}\label{equ: weight}
    w_{i} = \Gamma_{i}\times p(t_{{\rm E}, i}),
\end{equation}
where $\Gamma_i\equiv\theta_{{\rm E},i}\,\mu_{{\rm rel},i}$ represents the event rate, and $p(t_{{\rm E},i})$ is the likelihood terms for $t_{{\rm E},i}$ given the observational probability distributions of $t_{\mathrm{E,L1}} = 8.343 \pm 0.076$ day shown in Table \ref{tab:parm}.

The physical parameters inferred from the Bayesian analysis are illustrated in Figure~\ref{fig:baye}. These include the host mass $M_1$, the companion mass $M_2$, the lens distance $D_{\rm L}$ and the source distance $D_{\rm S}$ from Earth, the projected separation between the two lens components $a_\perp$, and the heliocentric lens–source relative proper motion $\mu_{\rm rel, hel}$. The host star is most consistent with a low-mass M dwarf located in the Galactic bulge, while the companion is likely a low-mass brown dwarf or a super-Jupiter. The projected separation is $\sim 450$ au. Although such a wide separation may raise concerns about long-term dynamical stability, similar systems have been identified previously. For example, the Wolf 940 system contains an M4 dwarf with a brown dwarf companion at a projected separation of $\sim 400$ au \citep{Wolf940}. Future detections of similar systems will enable population-level studies of such systems.

\section{Discussion}\label{dis}

\subsection{Prospects for Detecting Low-Mass Objects at Very Wide Orbits}

In this paper, we have presented the observations and analysis of the microlensing event \event, which was identified in a search for very wide-orbit binary-lens events among OGLE-IV and KMTNet data. We limited our search to these events because earlier surveys generally had lower observing cadences, making it difficult to distinguish bumps caused by microlensing from other variability based on light-curve morphology or color information. 

The upcoming Nancy Grace Roman Space Telescope \citep{Spergel2015_WFIRST,MatthewWFIRSTI} and the Earth 2.0 (ET) Microlensing Telescope \citep{CMST,ET} will be more sensitive to short-duration events, and thus to low-mass objects, thanks to their continuous and stable photometry. Their survey fields overlap with the primary fields of OGLE and KMTNet, enabling these space-based telescopes to detect additional low-mass objects on very wide orbits. Cross-matching Roman and ET events with historical ground-based detections will be crucial for constructing a complete census, as well as for verifying whether candidate FFPs are truly free-floating. Moreover, as OGLE and KMTNet continue their observations, additional similar events from their datasets are also anticipated.

An advantage of the Roman Space Telescope is that, when it detects the signature of a low-mass object, it may also resolve the host star through high-resolution imaging. Because the microlensing signal from the host occurs many years earlier and the host has since moved away from the source, Roman has a higher probability of resolving the host star in such events than in typical Roman planetary events. This enables a direct measurement of the lens-source relative proper motion, from which the angular Einstein radius can be derived via $\thetae = \te \times \mu_{\rm rel}$. Combined with photometric measurements of the host star, this information can be used to infer the lens mass and distance (e.g., \citealt{OB120950}).
For ET, its comparable photometric precision and overlapping observing window with KMTNet and OGLE enable simultaneous space- and ground-based observations to measure the satellite microlensing parallax \citep{1966MNRAS.134..315R,1994ApJ...421L..75G}. When combined with finite-source effects in the light curve, which are commonly present in planetary-induced bumps, this provides a way to determine the lens mass and distance \citep{KB240792}.

\subsection{Difficulties in Detecting Low-Mass Objects at Very Wide Orbits}

From the search for very wide-orbit binary-lens events and the analysis of \event, we identify three main difficulties in detecting such events. First, robustly distinguishing both bumps from other types of variability requires high-cadence, multi-band photometry. In the present case, the event occurred in the KMTNet and OGLE prime fields, which provided sufficient coverage in both $I$ and $V$ bands. By contrast, two other events that were initially identified as having short-lived bumps were ultimately shown to be cataclysmic variables.

Second, the 2L1S/1L2S degeneracy can present a significant challenge. The fundamental difference between the two models lies in the origin of the bumps: in a 2L1S model, both bumps arise from the same source star but correspond to two different $\te$, whereas in a 1L2S model, the two bumps are produced by two different source stars but share the same $\te$. To rule out the 1L2S model, both the timescale and source brightness of the two bumps must be well determined. In the present case, we are able to rule out the 1L2S model using this approach. However, it is generally rare for both bumps to be significantly magnified (i.e., small impact parameters). For a bump with a large impact parameter (e.g., $u_0 > 1$), the degeneracy among $(u_0, \te, f_{\rm S})$ may lead to poorly constrained timescale and source flux. In such cases, that bump could be fitted by a model sharing the same $\te$ as the other bump, making the 1L2S interpretation potentially viable.

In principle, the 1L2S model can be tested by the source colors, whereas the 2L1S model predicts achromatic variations \citep{Shude1991,Gaudi1998}. This test requires sufficiently dense coverage and high photometric precision in a secondary passband. However, there are cases in which the two sources have similar colors and both appear physically reasonable, rendering this diagnostic inconclusive. The present event is an example of such a case. From the 1L2S fit, the source properties are
\((V-I,I)_{\rm S1}=(2.47\pm0.06,\;18.336\pm0.015)\)~\HL{mag} and
\((V-I,I)_{\rm S2}=(2.45\pm0.07,\;20.623\pm0.041)\)~\HL{mag}. We place these sources on the CMD shown in Figure~\ref{fig:cmd}. The secondary source is about $0.6\,\sigma$ redder than the red boundary. Thus, although the 1L2S model would require an unusually metal-rich secondary star, for \event\ the 1L2S interpretation cannot be ruled out on the basis of the color information.

Third, it should be noted that the two widely separated bumps could also be produced by two unrelated 1L1S events. In this work, we present, to our knowledge for the first time, a detailed investigation of the 2$\times$1L1S model. The framework developed here to test the 2$\times$1L1S scenario, including source positions, total source flux, and source colors, can be applied in future studies. 

In summary, the identification of very wide-orbit binary-lens events requires both high-cadence, multi-band photometry and well-constrained light-curves to mitigate confusion with stellar variability and the 2L1S/1L2S/2$\times$1L1S degeneracy. Even with detailed color information, the degeneracy may remain unresolved in some cases.

\acknowledgments
H.Y. acknowledges support by the China Postdoctoral Science Foundation (No. 2024M762938). H.Y., W.Z., Y.T., J.Z., S.M., H.L. and Q.Q. acknowledge support by the National Natural Science Foundation of China (Grant No. 12133005). This work is part of the ET space mission which is funded by the China's Space Origins Exploration Program. This research has made use of the KMTNet system operated by the Korea Astronomy and Space Science Institute (KASI) at three host sites of CTIO in Chile, SAAO in South Africa, and SSO in Australia. Data transfer from the host site to KASI was supported by the Korea Research Environment Open NETwork (KREONET). This research was supported by KASI under the R\&D program (project No. 2026-1-904-01) supervised by the Ministry of Science and ICT. The OGLE project has received funding from the Polish National Science
Centre grant OPUS 2024/55/B/ST9/00447 awarded to A.U. This work is part of the ET space mission which is funded by the China's Space Origins Exploration Program. RP acknowledges support by the National Science Center, Poland, grant Sonata Bis 2021/42/E/ST9/00038. Work by J.C.Y. acknowledge support from N.S.F Grant No. AST-2108414. Work by C.H. was supported by the grants of National Research Foundation of Korea (2019R1A2C2085965 and 2020R1A4A2002885). Y.S. acknowledges support from BSF Grant No. 2020740.

\software{pySIS \citep{pysis,Yang_TLC,Yang_TLC2}, numpy \citep{numpy}, emcee \citep{emcee2,emcee}, Matplotlib \citep{Matplotlib}, SciPy \citep{scipy}, galpy \citep{bovy2015galpy}}

\clearpage

\def\appfloatletter{\ifnum\value{section}>0 \Alph{section}\else A\fi}
\setcounter{table}{0}
\setcounter{figure}{0}
\def\thetable{\appfloatletter\arabic{table}}
\def\thefigure{\appfloatletter\arabic{figure}}

\appendix

\section{Light curves for the two rejected candidates}
\label{rej-lc}
\begin{figure*}[h]
    \centering
    \includegraphics[width=\linewidth]{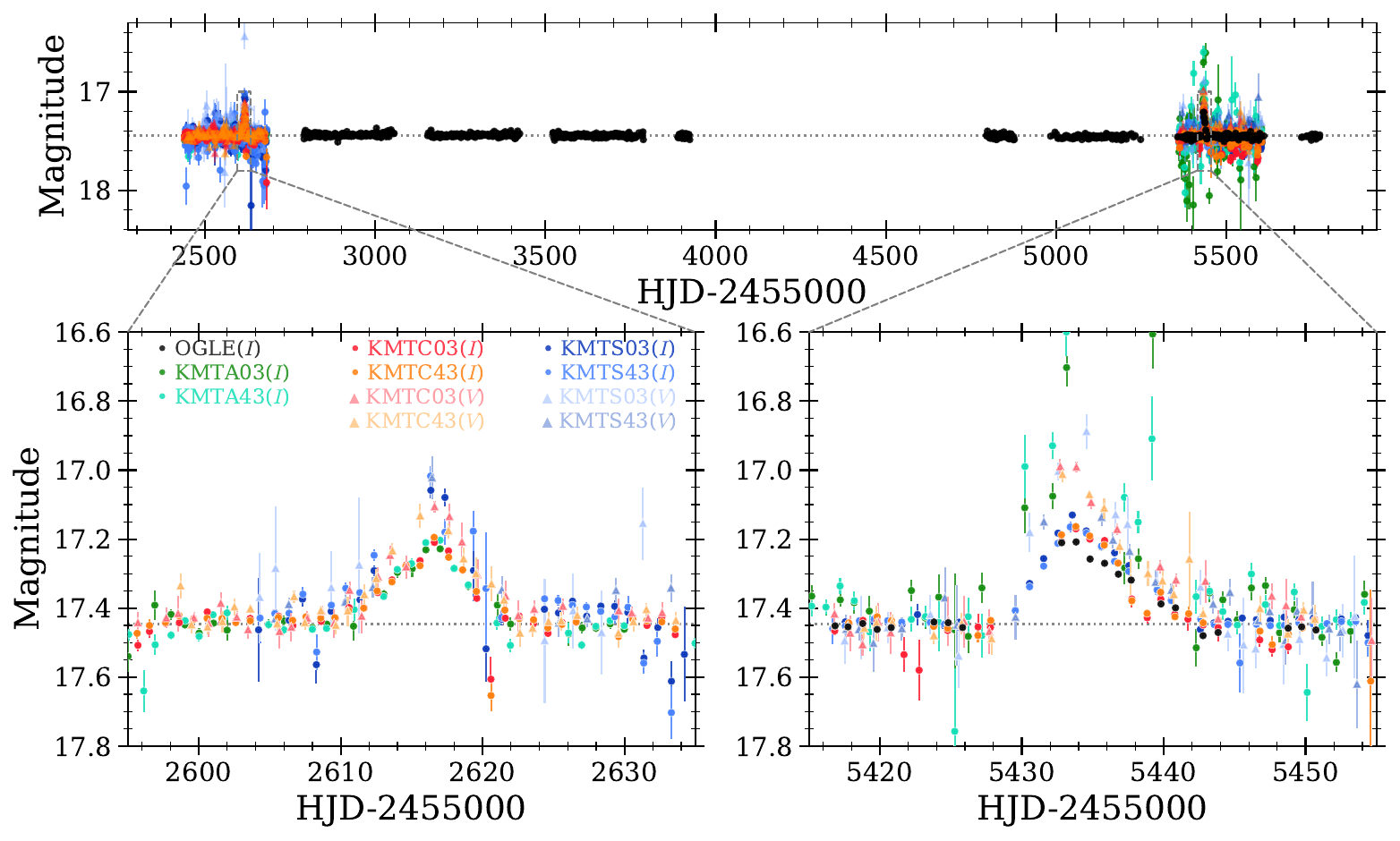}
    \caption{\HL{Light curve of the rejected candidate event, KMT-2016-BLG-0662/OGLE-2024-BLG-0548. Data from each dataset are binned daily for clarity.
    Only the KMTNet 2016 and 2024 data are re-reduced. OGLE data are from the OGLE-IV EWS \citep{Udalski1994, Udalski2003}. All datasets are aligned to the OGLE ($I$) observations by applying a constant magnitude offset determined from the baseline.}}
    \label{fig:lc}
\end{figure*}

\begin{figure*}[h]
    \centering
    \includegraphics[width=\linewidth]{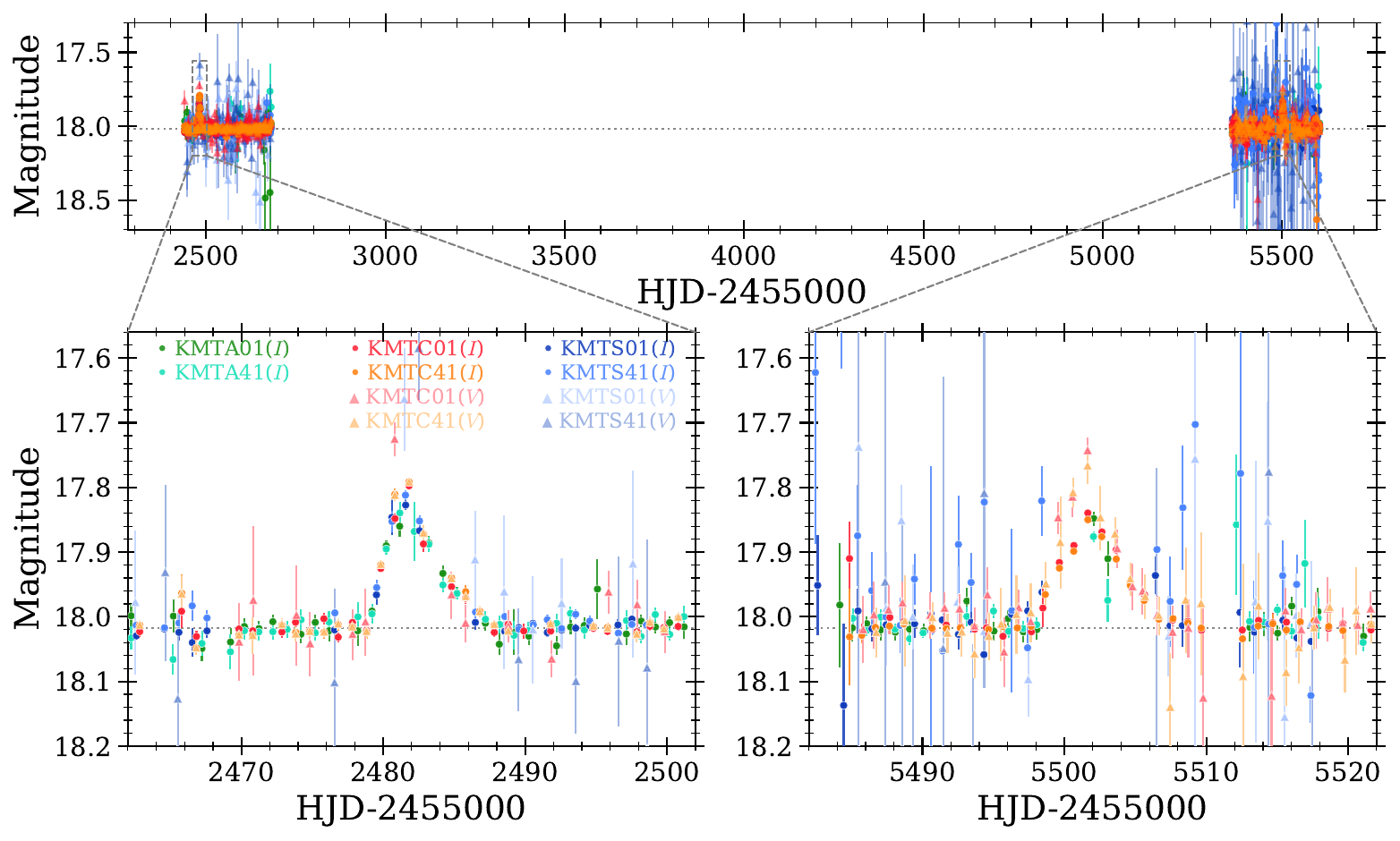}
    \caption{\HL{Light curve of the rejected candidate event, KMT-2016-BLG-1704/KMT-2024-BLG-1787. Data from each dataset are binned daily for clarity.
    Only the KMTNet 2016 and 2024 data are re-reduced. All datasets are aligned to the KMTC01 ($I$) observations by applying a constant magnitude offset determined from the baseline.}}
    \label{fig:lc}
\end{figure*}

\clearpage

\section{Dataset information of \event}
\begin{table*}[ht]
    \renewcommand\arraystretch{1.2}
    \centering
    \caption{Data information with corresponding error renormalization factors}
    \begin{threeparttable}
    \begin{tabular}{c c l c r c c c}
    \hline
    Collaboration & Site & Name & Filter & Coverage (${\rm HJD}^{\prime}$)\tnote{1}& $N_{\rm data}$ & $(k,\sigma_{\rm min})$\tnote{2} \\
    \hline
    OGLE & Las Campanas Observatory & OGLE & $I$ & 5260.8 -- 10932.6 & 13204 & $(1.55, 0.00)$  \\
    
    KMTNet & SSO & KMTA02-16 ($I$) & $I$ & 7440.3 -- 7679.9 & 1027 & $(1.23, 0.00)$ \\
    KMTNet & SSO & KMTA02-24 ($I$) & $I$ & 10361.2 -- 10604.9 & 2018 & $(1.10, 0.00)$ \\
    KMTNet & SSO & KMTA42-16 ($I$) & $I$ & 7443.3 -- 7679.9 & 1008 & $(1.09, 0.00)$ \\
    KMTNet & SSO & KMTA42-24 ($I$) & $I$ & 10361.2 -- 10604.9 & 2165 & $(1.18, 0.00)$ \\

    KMTNet & CTIO & KMTC02-16 ($I$) & $I$ & 7439.8 -- 7680.5 & 1458 & $(1.14, 0.00)$ \\
    KMTNet & CTIO & KMTC02-24 ($I$) & $I$ & 10349.9 -- 10604.5 & 2081 & $(0.84, 0.00)$ \\
    KMTNet & CTIO & KMTC42-16 ($I$) & $I$ & 7439.9 -- 7676.5 & 1330 & $(1.12, 0.00)$ \\
    KMTNet & CTIO & KMTC42-24 ($I$) & $I$ & 10347.9 -- 10604.5 & 2035 & $(0.87, 0.00)$ \\

    KMTNet & SAAO & KMTS02-16 ($I$) & $I$ & 7441.6 -- 7681.3 & 1570 & $(1.25, 0.00)$ \\
    KMTNet & SAAO & KMTS02-24 ($I$) & $I$ & 10361.6 -- 10604.3 & 1809 & $(1.01, 0.00)$ \\
    KMTNet & SAAO & KMTS42-16 ($I$) & $I$ & 7441.6 -- 7681.3 & 1670 & $(1.12, 0.00)$ \\
    KMTNet & SAAO & KMTS42-24 ($I$) & $I$ & 10361.6 -- 10604.2 & 2018 & $(1.21, 0.00)$ \\
           
    KMTNet & CTIO & KMTC02-16 ($V$) & $V$ & 7439.8 -- 7673.5 & 150 & $(0.96, 0.00)$ \\
    KMTNet & CTIO & KMTC02-24 ($V$) & $V$ & 10360.8 -- 10603.5 & 220 & $(1.00, 0.00)$ \\
    KMTNet & CTIO & KMTC42-16 ($V$) & $V$ & 7445.9 -- 7676.5 & 149 & $(1.09, 0.00)$ \\
    KMTNet & CTIO & KMTC42-24 ($V$) & $V$ & 10370.8 -- 10600.5 & 208 & $(1.20, 0.00)$ \\
    \hline
    \end{tabular}
   \label{tab:data}
   \begin{tablenotes}
   \item[1] ${\rm HJD}^{\prime} = {\rm HJD} - 2450000$.
    \item[2] $(k,\sigma_0)$ are the error renormalization factors as described in \cite{MB11293}.
    \end{tablenotes}
    \end{threeparttable}
\end{table*}

\section{Coefficients of the CMD red boundary}
\begin{table}[h]
    \centering
    \caption{\HL{Coefficients of Equation~\ref{equ:VIM}}}
    \label{tab:VIM}
    \begin{tabular}{cc}
        \hline
        $i$ & $a_i$ \\
        \hline
        0 & $2.3$ \\
        1 & $6.9\times10^{-1}$ \\
        2 & $-1.1\times10^{-1}$ \\
        3 & $-6.6\times10^{-2}$ \\
        4 & $1.9\times10^{-2}$ \\
        5 & $7.3\times10^{-3}$ \\
        6 & $-8.9\times10^{-4}$ \\
        7 & $-3.0\times10^{-4}$ \\
        \hline
    \end{tabular}
\end{table}

\section{Gaia Proper Motions}

\begin{figure}[htb] 
    \centering
    \includegraphics[width=0.45\textwidth]{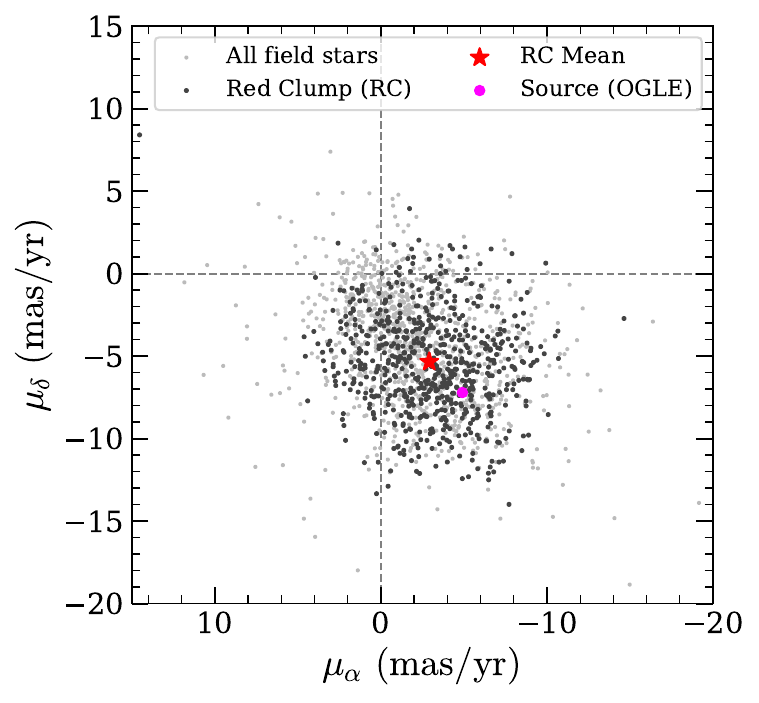}
    \caption{The proper motions of the source star and field stars within a $1.5'$ radius are shown. The proper motions of the field stars are measured using \textit{Gaia} DR3. The source star, indicated by the orange dot, has no proper-motion measurement in \textit{Gaia} DR3, and its value is derived from the OGLE-URANUS astrometry database (Udalski et al., in prep.). Red clump (RC) stars are shown as black dots, and their mean proper motion is indicated by the red dot. \HL{The RC mean proper motion is $(\mu_\alpha,\mu_\delta)=(-2.91\pm0.12,-5.34\pm0.12)~{\rm mas/yr}$, and the source proper motion is $(\mu_\alpha,\mu_\delta)=(-4.91\pm0.24,-7.21\pm0.17)~{\rm mas/yr}$.}}
    \label{fig:pm}
\end{figure}

To determine the heliocentric proper motion of the reference field stars used for astrometric calibration, we extract proper motions from \textit{Gaia} DR3 \citep{GaiaDR3} for field stars within a $1.5'$ radius centered on the event. Red clump stars are selected using $2.1 < G_{BP} - G_{RP} < 3.1$ and $15.0 < G_{RP} < 16.7$. Their mean proper motion is $\mu_{\rm RC} ({\rm N, E}) = (-5.34 \pm 0.12, -2.91 \pm 0.12)~{\rm mas~yr^{-1}}$, corresponding to $\mu_{\rm RC} ({\ell, b}) = (-6.06 \pm 0.12, -0.09 \pm 0.12)~{\rm mas~yr^{-1}}$ in Galactic coordinates, which is adopted as the astrometric reference frame in Section~\ref{2-1L1S}. The corresponding proper-motion distribution, together with the microlensing source proper motion, is shown in Figure~\ref{fig:pm}. The source has a proper motion typical of bulge stars.

\bibliography{Zang.bib}

@ARTICLE{bovy2015galpy,
       author = {{Bovy}, Jo},
        title = "{galpy: A python Library for Galactic Dynamics}",
      journal = {\apjs},
         year = 2015,
        month = feb,
       volume = {216},
       number = {2},
          eid = {29},
        pages = {29},
          doi = {10.1088/0067-0049/216/2/29},
archivePrefix = {arXiv},
       eprint = {1412.3451},
 primaryClass = {astro-ph.GA},
       adsurl = {https://ui.adsabs.harvard.edu/abs/2015ApJS..216...29B}
}

@ARTICLE{OGLE_wide_binary,
       author = {{Oliveira}, R.~A.~P. and {Poleski}, R. and {Mr{\'o}z}, P. and {Udalski}, A. and {Skowron}, J. and {Mr{\'o}z}, M.~J. and {Szyma{\'n}ski}, M.~K. and {Soszy{\'n}ski}, I. and {Ulaczyk}, K. and {Pietrukowicz}, P. and {Rybicki}, K.~A. and {Iwanek}, P. and {Wrona}, M. and {Gromadzki}, M.},
        title = "{Automated Detection and Modeling of Binary Microlensing Events in OGLE-IV data. I. Events with Well-Separated Bumps}",
      journal = {\actaa},
         year = 2025,
        month = aug,
       volume = {75},
       number = {1},
        pages = {27-52},
          doi = {10.32023/0001-5237/75.1.2},
archivePrefix = {arXiv},
       eprint = {2504.21085},
 primaryClass = {astro-ph.GA},
       adsurl = {https://ui.adsabs.harvard.edu/abs/2025AcA....75...27O}
}

@ARTICLE{Spergel2015_WFIRST,
       author = {{Spergel}, D. and {Gehrels}, N. and {Baltay}, C. and {Bennett}, D. and {Breckinridge}, J. and {Donahue}, M. and {Dressler}, A. and {Gaudi}, B.~S. and {Greene}, T. and {Guyon}, O. and {Hirata}, C. and {Kalirai}, J. and {Kasdin}, N.~J. and {Macintosh}, B. and {Moos}, W. and {Perlmutter}, S. and {Postman}, M. and {Rauscher}, B. and {Rhodes}, J. and {Wang}, Y. and {Weinberg}, D. and {Benford}, D. and {Hudson}, M. and {Jeong}, W. -S. and {Mellier}, Y. and {Traub}, W. and {Yamada}, T. and {Capak}, P. and {Colbert}, J. and {Masters}, D. and {Penny}, M. and {Savransky}, D. and {Stern}, D. and {Zimmerman}, N. and {Barry}, R. and {Bartusek}, L. and {Carpenter}, K. and {Cheng}, E. and {Content}, D. and {Dekens}, F. and {Demers}, R. and {Grady}, K. and {Jackson}, C. and {Kuan}, G. and {Kruk}, J. and {Melton}, M. and {Nemati}, B. and {Parvin}, B. and {Poberezhskiy}, I. and {Peddie}, C. and {Ruffa}, J. and {Wallace}, J.~K. and {Whipple}, A. and {Wollack}, E. and {Zhao}, F.},
        title = "{Wide-Field InfrarRed Survey Telescope-Astrophysics Focused Telescope Assets WFIRST-AFTA 2015 Report}",
      journal = {arXiv e-prints},
         year = 2015,
        month = mar,
          eid = {arXiv:1503.03757},
        pages = {arXiv:1503.03757},
archivePrefix = {arXiv},
       eprint = {1503.03757},
 primaryClass = {astro-ph.IM},
       adsurl = {https://ui.adsabs.harvard.edu/abs/2015arXiv150303757S}
}

@ARTICLE{NASAExo,
       author = {{Christiansen}, Jessie L. and {McElroy}, Douglas L. and {Harbut}, Marcy and {Ciardi}, David R. and {Crane}, Megan and {Good}, John and {Hardegree-Ullman}, Kevin K. and {Kesseli}, Aurora Y. and {Lund}, Michael B. and {Lynn}, Meca and {Muthiar}, Ananda and {Nilsson}, Ricky and {Oluyide}, Toba and {Papin}, Michael and {Rivera}, Amalia and {Swain}, Melanie and {Susemiehl}, Nicholas D. and {Tam}, Raymond and {van Eyken}, Julian and {Beichman}, Charles},
        title = "{The NASA Exoplanet Archive and Exoplanet Follow-up Observing Program: Data, Tools, and Usage}",
      journal = {PSJ},
         year = 2025,
        month = aug,
       volume = {6},
       number = {8},
          eid = {186},
        pages = {186},
          doi = {10.3847/PSJ/ade3c2},
archivePrefix = {arXiv},
       eprint = {2506.03299},
 primaryClass = {astro-ph.EP},
       adsurl = {https://ui.adsabs.harvard.edu/abs/2025PSJ.....6..186C}
}

@ARTICLE{OB160007,
       author = {{Zang}, Weicheng and {Jung}, Youn Kil and {Yee}, Jennifer C. and {Hwang}, Kyu-Ha and {Yang}, Hongjing and {Udalski}, Andrzej and {Sumi}, Takahiro and {Gould}, Andrew and {Mao}, Shude and {Albrow}, Michael D. and {Chung}, Sun-Ju and {Han}, Cheongho and {Ryu}, Yoon-Hyun and {Shin}, In-Gu and {Shvartzvald}, Yossi and {Cha}, Sang-Mok and {Kim}, Dong-Jin and {Kim}, Hyoun-Woo and {Kim}, Seung-Lee and {Lee}, Chung-Uk and {Lee}, Dong-Joo and {Lee}, Yongseok and {Park}, Byeong-Gon and {Pogge}, Richard W. and {Zhang}, Xiangyu and {Kuang}, Renkun and {Wang}, Hanyue and {Zhang}, Jiyuan and {Hu}, Zhecheng and {Zhu}, Wei and {Mr{\'o}z}, Przemek and {Skowron}, Jan and {Poleski}, Rados{\l}aw and {Szyma{\'n}ski}, Micha{\l} K. and {Soszy{\'n}ski}, Igor and {Pietrukowicz}, Pawe{\l} and {Koz{\l}owski}, Szymon and {Ulaczyk}, Krzysztof and {Rybicki}, Krzysztof A. and {Iwanek}, Patryk and {Wrona}, Marcin and {Gromadzki}, Mariusz and {Abe}, Fumio and {Barry}, Richard and {Bennett}, David P. and {Bhattacharya}, Aparna and {Bond}, Ian A. and {Fujii}, Hirosane and {Fukui}, Akihiko and {Hamada}, Ryusei and {Hirao}, Yuki and {Silva}, Stela Ishitani and {Itow}, Yoshitaka and {Kirikawa}, Rintaro and {Koshimoto}, Naoki and {Matsubara}, Yutaka and {Miyazaki}, Shota and {Muraki}, Yasushi and {Olmschenk}, Greg and {Ranc}, Cl{\'e}ment and {Rattenbury}, Nicholas J. and {Satoh}, Yuki and {Suzuki}, Daisuke and {Tomoyoshi}, Mio and {Tristram}, Paul J. and {Vandorou}, Aikaterini and {Yama}, Hibiki and {Yamashita}, Kansuke},
        title = "{Microlensing events indicate that super-Earth exoplanets are common in Jupiter-like orbits}",
      journal = {Science},
         year = 2025,
        month = apr,
       volume = {388},
       number = {6745},
        pages = {400-404},
          doi = {10.1126/science.adn6088},
archivePrefix = {arXiv},
       eprint = {2504.20158},
 primaryClass = {astro-ph.EP},
       adsurl = {https://ui.adsabs.harvard.edu/abs/2025Sci...388..400Z}
}

@ARTICLE{Yang2021_GalacticModel,
       author = {{Yang}, Hongjing and {Mao}, Shude and {Zang}, Weicheng and {Zhang}, Xiangyu},
        title = "{Microlensing predictions: impact of Galactic disc dynamical models}",
      journal = {\mnras},
         year = 2021,
        month = apr,
       volume = {502},
       number = {4},
        pages = {5631-5642},
          doi = {10.1093/mnras/stab441},
archivePrefix = {arXiv},
       eprint = {2010.16146},
 primaryClass = {astro-ph.GA},
       adsurl = {https://ui.adsabs.harvard.edu/abs/2021MNRAS.502.5631Y}
}

@Article{Matplotlib,
  Author    = {Hunter, J. D.},
  Title     = {Matplotlib: A 2D graphics environment},
  Journal   = {Computing in Science \& Engineering},
  Volume    = {9},
  Number    = {3},
  Pages     = {90--95},
  publisher = {IEEE COMPUTER SOC},
  doi       = {10.1109/MCSE.2007.55},
  year      = 2007
}

@Article{numpy,
 title         = {Array programming with {NumPy}},
 author        = {Charles R. Harris and K. Jarrod Millman and St{\'{e}}fan J.
                 van der Walt and Ralf Gommers and Pauli Virtanen and David
                 Cournapeau and Eric Wieser and Julian Taylor and Sebastian
                 Berg and Nathaniel J. Smith and Robert Kern and Matti Picus
                 and Stephan Hoyer and Marten H. van Kerkwijk and Matthew
                 Brett and Allan Haldane and Jaime Fern{\'{a}}ndez del
                 R{\'{i}}o and Mark Wiebe and Pearu Peterson and Pierre
                 G{\'{e}}rard-Marchant and Kevin Sheppard and Tyler Reddy and
                 Warren Weckesser and Hameer Abbasi and Christoph Gohlke and
                 Travis E. Oliphant},
 year          = {2020},
 month         = sep,
 journal       = {Nature},
 volume        = {585},
 number        = {7825},
 pages         = {357--362},
 doi           = {10.1038/s41586-020-2649-2},
 publisher     = {Springer Science and Business Media {LLC}},
 url           = {https://doi.org/10.1038/s41586-020-2649-2}
}

@ARTICLE{Nemiroff1994,
   author = {{Nemiroff}, R.~J. and {Wickramasinghe}, W.~A.~D.~T.},
    title = "{Finite source sizes and the information content of macho-type lens search light curves}",
  journal = {\apjl},
   eprint = {astro-ph/9401005},
     year = 1994,
    month = mar,
   volume = 424,
    pages = {L21-L23},
      doi = {10.1086/187265},
   adsurl = {http://adsabs.harvard.edu/abs/1994ApJ...424L..21N}
}

@ARTICLE{Gould2000,
   author = {{Gould}, A.},
    title = "{A Natural Formalism for Microlensing}",
  journal = {\apj},
   eprint = {astro-ph/0001421},
     year = 2000,
    month = oct,
   volume = 542,
    pages = {785-788},
      doi = {10.1086/317037},
   adsurl = {http://adsabs.harvard.edu/abs/2000ApJ...542..785G}
}

@ARTICLE{Sumi2011,
   author = {{Sumi}, T. and {Kamiya}, K. and {Bennett}, D.~P. and {Bond}, I.~A. and 
	{Abe}, F. and {Botzler}, C.~S. and {Fukui}, A. and {Furusawa}, K. and 
	{Hearnshaw}, J.~B. and {Itow}, Y. and {Kilmartin}, P.~M. and 
	{Korpela}, A. and {Lin}, W. and {Ling}, C.~H. and {Masuda}, K. and 
	{Matsubara}, Y. and {Miyake}, N. and {Motomura}, M. and {Muraki}, Y. and 
	{Nagaya}, M. and {Nakamura}, S. and {Ohnishi}, K. and {Okumura}, T. and 
	{Perrott}, Y.~C. and {Rattenbury}, N. and {Saito}, T. and {Sako}, T. and 
	{Sullivan}, D.~J. and {Sweatman}, W.~L. and {Tristram}, P.~J. and 
	{}, A. and {Szyma{\'n}ski}, M.~K. and {Kubiak}, M. and 
	{Pietrzy{\'n}ski}, G. and {Poleski}, R. and {Soszy{\'n}ski}, I. and 
	{Wyrzykowski}, {\L}. and {Ulaczyk}, K. and {Microlensing Observations in Astrophysics (MOA) Collaboration}
	},
    title = "{Unbound or distant planetary mass population detected by gravitational microlensing}",
  journal = {\nat},
archivePrefix = "arXiv",
   eprint = {1105.3544},
 primaryClass = "astro-ph.EP",
     year = 2011,
    month = may,
   volume = 473,
    pages = {349-352},
      doi = {10.1038/nature10092},
   adsurl = {http://adsabs.harvard.edu/abs/2011Natur.473..349S}
}

@ARTICLE{Nataf2013,
   author = {{Nataf}, D.~M. and {Gould}, A. and {Fouqu{\'e}}, P. and {Gonzalez}, O.~A. and 
	{Johnson}, J.~A. and {Skowron}, J. and {}, A. and {Szyma{\'n}ski}, M.~K. and 
	{Kubiak}, M. and {Pietrzy{\'n}ski}, G. and {Soszy{\'n}ski}, I. and 
	{Ulaczyk}, K. and {Wyrzykowski}, {\L}. and {Poleski}, R.},
    title = "{Reddening and Extinction toward the Galactic Bulge from OGLE-III: The Inner Milky Way's R$_{V}$ \~{} 2.5 Extinction Curve}",
  journal = {\apj},
archivePrefix = "arXiv",
   eprint = {1208.1263},
     year = 2013,
    month = jun,
   volume = 769,
      eid = {88},
    pages = {88},
      doi = {10.1088/0004-637X/769/2/88},
   adsurl = {http://adsabs.harvard.edu/abs/2013ApJ...769...88N}
}

@ARTICLE{Mroz2017a,
   author = {{Mr{\'o}z}, P. and {}, A. and {Skowron}, J. and {Poleski}, R. and 
	{Koz{\l}owski}, S. and {Szyma{\'n}ski}, M.~K. and {Soszy{\'n}ski}, I. and 
	{Wyrzykowski}, {\L}. and {Pietrukowicz}, P. and {Ulaczyk}, K. and 
	{Skowron}, D. and {Pawlak}, M.},
    title = "{No large population of unbound or wide-orbit Jupiter-mass planets}",
  journal = {\nat},
archivePrefix = "arXiv",
   eprint = {1707.07634},
 primaryClass = "astro-ph.EP",
     year = 2017,
    month = aug,
   volume = 548,
    pages = {183-186},
      doi = {10.1038/nature23276},
   adsurl = {http://adsabs.harvard.edu/abs/2017Natur.548..183M}
}

@ARTICLE{1994ApJ...421L..75G,
   author = {{Gould}, A.},
    title = "{MACHO velocities from satellite-based parallaxes}",
  journal = {\apjl},
     year = 1994,
    month = feb,
   volume = 421,
    pages = {L75-L78},
      doi = {10.1086/187191},
   adsurl = {http://adsabs.harvard.edu/abs/1994ApJ...421L..75G}
}

@ARTICLE{Zhu2017spitzer,
   author = {{Zhu}, W. and {Udalski}, A. and {Calchi Novati}, S. and {Chung}, S.-J. and 
	{Jung}, Y.~K. and {Ryu}, Y.-H. and {}, I.-G. and {Gould}, A. and 
	{Lee}, C.-U. and {Albrow}, M.~D. and {Yee}, J.~C. and {Han}, C. and 
	{Hwang}, K.-H. and {Cha}, S.-M. and {Kim}, D.-J. and {Kim}, H.-W. and 
	{Kim}, S.-L. and {Kim}, Y.-H. and {Lee}, Y. and {Park}, B.-G. and 
	{Pogge}, R.~W. and {KMTNet Collaboration} and {Poleski}, R. and 
	{Mr{\'o}z}, P. and {Pietrukowicz}, P. and {Skowron}, J. and 
	{Szyma{\'n}ski}, M.~K. and {Koz{\l}owski}, S. and {Ulaczyk}, K. and 
	{Pawlak}, M. and {OGLE Collaboration} and {Beichman}, C. and 
	{Bryden}, G. and {Carey}, S. and {Fausnaugh}, M. and {Gaudi}, B.~S. and 
	{Henderson}, C.~B. and {Shvartzvald}, Y. and {Wibking}, B. and 
	{Spitzer Team}},
    title = "{Toward a Galactic Distribution of Planets. I. Methodology and Planet Sensitivities of the 2015 High-cadence Spitzer Microlens Sample}",
  journal = {\aj},
archivePrefix = "arXiv",
   eprint = {1701.05191},
 primaryClass = "astro-ph.EP",
     year = 2017,
    month = nov,
   volume = 154,
      eid = {210},
    pages = {210},
      doi = {10.3847/1538-3881/aa8ef1},
   adsurl = {http://adsabs.harvard.edu/abs/2017AJ....154..210Z}
}

@ARTICLE{Alard1998,
   author = {{Alard}, C. and {Lupton}, R.~H.},
    title = "{A Method for Optimal Image Subtraction}",
  journal = {\apj},
   eprint = {astro-ph/9712287},
     year = 1998,
    month = aug,
   volume = 503,
    pages = {325-331},
      doi = {10.1086/305984},
   adsurl = {http://adsabs.harvard.edu/abs/1998ApJ...503..325A}
}

@ARTICLE{Udalski2003,
   author = {{Udalski}, A.},
    title = "{The Optical Gravitational Lensing Experiment. Real Time Data Analysis Systems in the OGLE-III Survey}",
  journal = {\actaa},
   eprint = {astro-ph/0401123},
     year = 2003,
    month = dec,
   volume = 53,
    pages = {291-305},
   adsurl = {http://adsabs.harvard.edu/abs/2003AcA....53..291U}
}

@ARTICLE{emcee,
   author = {{Foreman-Mackey}, D. and {Hogg}, D.~W. and {Lang}, D. and {Goodman}, J.
	},
    title = "{emcee: The MCMC Hammer}",
  journal = {\pasp},
archivePrefix = "arXiv",
   eprint = {1202.3665},
 primaryClass = "astro-ph.IM",
     year = 2013,
    month = mar,
   volume = 125,
    pages = {306},
      doi = {10.1086/670067},
   adsurl = {http://adsabs.harvard.edu/abs/2013PASP..125..306F}
}

@ARTICLE{OGLEIII,
   author = {{Szyma{\'n}ski}, M.~K. and {Udalski}, A. and {Soszy{\'n}ski}, I. and 
	{Kubiak}, M. and {Pietrzy{\'n}ski}, G. and {Poleski}, R. and 
	{Wyrzykowski}, {\L}. and {Ulaczyk}, K.},
    title = "{The Optical Gravitational Lensing Experiment. OGLE-III Photometric Maps of the Galactic Bulge Fields}",
  journal = {\actaa},
archivePrefix = "arXiv",
   eprint = {1107.4008},
 primaryClass = "astro-ph.SR",
     year = 2011,
    month = jun,
   volume = 61,
    pages = {83-102},
   adsurl = {http://adsabs.harvard.edu/abs/2011AcA....61...83S}
}

@ARTICLE{Udalski1994,
   author = {{Udalski}, A. and {Szymanski}, M. and {Kaluzny}, J. and {Kubiak}, M. and 
	{Mateo}, M. and {Krzeminski}, W. and {Paczynski}, B.},
    title = "{The Optical Gravitational Lensing Experiment. The Early Warning System: Real Time Microlensing}",
  journal = {\actaa},
   eprint = {astro-ph/9408026},
     year = 1994,
    month = jul,
   volume = 44,
    pages = {227-234},
   adsurl = {http://adsabs.harvard.edu/abs/1994AcA....44..227U}
}

@ARTICLE{1966MNRAS.134..315R,
   author = {{Refsdal}, S.},
    title = "{On the possibility of determining the distances and masses of stars from the gravitational lens effect}",
  journal = {\mnras},
     year = 1966,
   volume = 134,
    pages = {315},
      doi = {10.1093/mnras/134.3.315},
   adsurl = {http://adsabs.harvard.edu/abs/1966MNRAS.134..315R}
}

@ARTICLE{KMT2016,
   author = {{Kim}, S.-L. and {Lee}, C.-U. and {Park}, B.-G. and {Kim}, D.-J. and 
	{Cha}, S.-M. and {Lee}, Y. and {Han}, C. and {Chun}, M.-Y. and 
	{Yuk}, I.},
    title = "{KMTNET: A Network of 1.6 m Wide-Field Optical Telescopes Installed at Three Southern Observatories}",
  journal = {Journal of Korean Astronomical Society},
     year = 2016,
    month = feb,
   volume = 49,
    pages = {37-44},
      doi = {10.5303/JKAS.2016.49.1.37},
   adsurl = {http://adsabs.harvard.edu/abs/2016JKAS...49...37K}
}

@ARTICLE{Wolf940,
       author = {{Leggett}, S.~K. and {Saumon}, D. and {Burningham}, Ben and {Cushing}, Michael C. and {Marley}, M.~S. and {Pinfield}, D.~J.},
        title = "{Properties of the T8.5 Dwarf Wolf 940 B}",
      journal = {\apj},
         year = 2010,
        month = sep,
       volume = {720},
       number = {1},
        pages = {252-258},
          doi = {10.1088/0004-637X/720/1/252},
archivePrefix = {arXiv},
       eprint = {1007.1252},
 primaryClass = {astro-ph.SR},
       adsurl = {https://ui.adsabs.harvard.edu/abs/2010ApJ...720..252L}
}

@ARTICLE{Shude1994,
   author = {{Witt}, H.~J. and {Mao}, S.},
    title = "{Can lensed stars be regarded as pointlike for microlensing by MACHOs?}",
  journal = {\apj},
     year = 1994,
    month = aug,
   volume = 430,
    pages = {505-510},
      doi = {10.1086/174426},
   adsurl = {http://adsabs.harvard.edu/abs/1994ApJ...430..505W}
}

@ARTICLE{Shude1991,
   author = {{Mao}, S. and {Paczynski}, B.},
    title = "{Gravitational microlensing by double stars and planetary systems}",
  journal = {\apjl},
     year = 1991,
    month = jun,
   volume = 374,
    pages = {L37-L40},
      doi = {10.1086/186066},
   adsurl = {http://adsabs.harvard.edu/abs/1991ApJ...374L..37M}
}

@ARTICLE{Andy1992,
   author = {{Gould}, A. and {Loeb}, A.},
    title = "{Discovering planetary systems through gravitational microlenses}",
  journal = {\apj},
     year = 1992,
    month = sep,
   volume = 396,
    pages = {104-114},
      doi = {10.1086/171700},
   adsurl = {http://adsabs.harvard.edu/abs/1992ApJ...396..104G}
}

@ARTICLE{OGLEIV,
   author = {{Udalski}, A. and {Szyma{\'n}ski}, M.~K. and {Szyma{\'n}ski}, G.
	},
    title = "{OGLE-IV: Fourth Phase of the Optical Gravitational Lensing Experiment}",
  journal = {\actaa},
archivePrefix = "arXiv",
   eprint = {1504.05966},
 primaryClass = "astro-ph.SR",
     year = 2015,
    month = mar,
   volume = 65,
    pages = {1-38},
   adsurl = {http://adsabs.harvard.edu/abs/2015AcA....65....1U}
}

@ARTICLE{Gaudi1998,
   author = {{Gaudi}, B.~S.},
    title = "{Distinguig Between Binary-Source and Planetary Microlensing Perturbations}",
  journal = {\apj},
     year = 1998,
    month = oct,
   volume = 506,
    pages = {533-539},
      doi = {10.1086/306256},
   adsurl = {http://adsabs.harvard.edu/abs/1998ApJ...506..533G}
}

@ARTICLE{pysis,
   author = {{Albrow}, M.~D. and {Horne}, K. and {Bramich}, D.~M. and {Fouqu{\'e}}, P. and 
	{Miller}, V.~R. and {Beaulieu}, J.-P. and {Coutures}, C. and 
	{Menzies}, J. and {Williams}, A. and {Batista}, V. and {Bennett}, D.~P. and 
	{Brillant}, S. and {Cassan}, A. and {Dieters}, S. and {Dominis Prester}, D. and 
	{Donatowicz}, J. and {Greenhill}, J. and {Kains}, N. and {Kane}, S.~R. and 
	{Kubas}, D. and {Marquette}, J.~B. and {Pollard}, K.~R. and 
	{Sahu}, K.~C. and {Tsapras}, Y. and {Wambsganss}, J. and {Zub}, M.
	},
    title = "{Difference imaging photometry of blended gravitational microlensing events with a numerical kernel}",
  journal = {\mnras},
archivePrefix = "arXiv",
   eprint = {0905.3003},
 primaryClass = "astro-ph.SR",
     year = 2009,
    month = aug,
   volume = 397,
    pages = {2099-2105},
      doi = {10.1111/j.1365-2966.2009.15098.x},
   adsurl = {http://adsabs.harvard.edu/abs/2009MNRAS.397.2099A}
}

@ARTICLE{MB12486,
   author = {{Hwang}, K.-H. and {Choi}, J.-Y. and {Bond}, I.~A. and {Sumi}, T. and 
	{Han}, C. and {Gaudi}, B.~S. and {Gould}, A. and {Bozza}, V. and 
	{Beaulieu}, J.-P. and {Tsapras}, Y. and {Abe}, F. and {Bennett}, D.~P. and 
	{Botzler}, C.~S. and {Chote}, P. and {Freeman}, M. and {Fukui}, A. and 
	{Fukunaga}, D. and {Harris}, P. and {Itow}, Y. and {Koshimoto}, N. and 
	{Ling}, C.~H. and {Masuda}, K. and {Matsubara}, Y. and {Muraki}, Y. and 
	{Namba}, S. and {Ohnishi}, K. and {Rattenbury}, N.~J. and {Saito}, T. and 
	{Sullivan}, D.~J. and {Sweatman}, W.~L. and {Suzuki}, D. and 
	{Tristram}, P.~J. and {Wada}, K. and {Yamai}, N. and {Yock}, P.~C.~M. and 
	{Yonehara}, A. and {MOA Collaboration} and {de Almeida}, L.~A. and 
	{DePoy}, D.~L. and {Dong}, S. and {Jablonski}, F. and {Jung}, Y.~K. and 
	{Kavka}, A. and {Lee}, C.-U. and {Park}, H. and {Pogge}, R.~W. and 
	{}, I.-G. and {Yee}, J.~C. and {{$\mu$}FUN Collaboration} and 
	{Albrow}, M.~D. and {Bachelet}, E. and {Batista}, V. and {Brillant}, S. and 
	{Caldwell}, J.~A.~R. and {Cassan}, A. and {Cole}, A. and {Corrales}, E. and 
	{Coutures}, C. and {Dieters}, S. and {Dominis Prester}, D. and 
	{Donatowicz}, J. and {Fouqu{\'e}}, P. and {Greenhill}, J. and 
	{J{\o}rgensen}, U.~G. and {Kane}, S.~R. and {Kubas}, D. and 
	{Marquette}, J.-B. and {Martin}, R. and {Meintjes}, P. and {Menzies}, J. and 
	{Pollard}, K.~R. and {Williams}, A. and {Wouters}, D. and {PLANET Collaboration} and 
	{Bramich}, D.~M. and {Dominik}, M. and {Horne}, K. and {Browne}, P. and 
	{Hundertmark}, M. and {Ipatov}, S. and {Kains}, N. and {Snodgrass}, C. and 
	{Steele}, I.~A. and {Street}, R.~A. and {RoboNet Collaboration}
	},
    title = "{Interpretation of a Short-term Anomaly in the Gravitational Microlensing Event MOA-2012-BLG-486}",
  journal = {\apj},
archivePrefix = "arXiv",
   eprint = {1308.5762},
 primaryClass = "astro-ph.SR",
     year = 2013,
    month = nov,
   volume = 778,
      eid = {55},
    pages = {55},
      doi = {10.1088/0004-637X/778/1/55},
   adsurl = {http://adsabs.harvard.edu/abs/2013ApJ...778...55H}
}

@ARTICLE{Yoo2004,
   author = {{Yoo}, J. and {DePoy}, D.~L. and {Gal-Yam}, A. and {Gaudi}, B.~S. and 
	{Gould}, A. and {Han}, C. and {Lipkin}, Y. and {Maoz}, D. and 
	{Ofek}, E.~O. and {Park}, B.-G. and {Pogge}, R.~W. and {Mu-Fun Collaboration} and 
	{Udalski}, A. and {Soszy{\'n}ski}, I. and {Wyrzykowski}, {\L}. and 
	{Kubiak}, M. and {Szyma{\'n}ski}, M. and {Pietrzy{\'n}ski}, G. and 
	{Szewczyk}, O. and {{\.Z}ebru{\'n}}, K. and {OGLE Collaboration}
	},
    title = "{OGLE-2003-BLG-262: Finite-Source Effects from a Point-Mass Lens}",
  journal = {\apj},
   eprint = {astro-ph/0309302},
     year = 2004,
    month = mar,
   volume = 603,
    pages = {139-151},
      doi = {10.1086/381241},
   adsurl = {http://adsabs.harvard.edu/abs/2004ApJ...603..139Y}
}

@ARTICLE{Wozniak2000,
   author = {{Wozniak}, P.~R.},
    title = "{Difference Image Analysis of the OGLE-II Bulge Data. I. The Method}",
  journal = {\actaa},
   eprint = {astro-ph/0012143},
     year = 2000,
    month = dec,
   volume = 50,
    pages = {421-450},
   adsurl = {http://adsabs.harvard.edu/abs/2000AcA....50..421W}
}

@ARTICLE{Sumi2023,
       author = {{Sumi}, Takahiro and {Koshimoto}, Naoki and {Bennett}, David P. and {Rattenbury}, Nicholas J. and {Abe}, Fumio and {Barry}, Richard and {Bhattacharya}, Aparna and {Bond}, Ian A. and {Fujii}, Hirosane and {Fukui}, Akihiko and {Hamada}, Ryusei and {Hirao}, Yuki and {Silva}, Stela Ishitani and {Itow}, Yoshitaka and {Kirikawa}, Rintaro and {Kondo}, Iona and {Matsubara}, Yutaka and {Miyazaki}, Shota and {Muraki}, Yasushi and {Olmschenk}, Greg and {Ranc}, Cl{\'e}ment and {Satoh}, Yuki and {Suzuki}, Daisuke and {Tomoyoshi}, Mio and {Tristram}, Paul. J. and {Vandorou}, Aikaterini and {Yama}, Hibiki and {Yamashita}, Kansuke},
        title = "{Free-floating Planet Mass Function from MOA-II 9 yr Survey toward the Galactic Bulge}",
      journal = {\aj},
         year = 2023,
        month = sep,
       volume = {166},
       number = {3},
          eid = {108},
        pages = {108},
          doi = {10.3847/1538-3881/ace688},
archivePrefix = {arXiv},
       eprint = {2303.08280},
 primaryClass = {astro-ph.EP},
       adsurl = {https://ui.adsabs.harvard.edu/abs/2023AJ....166..108S}
}

@ARTICLE{Gould2022_FFP_EinsteinDesert,
       author = {{Gould}, Andrew and {Jung}, Youn Kil and {Hwang}, Kyu-Ha and {Dong}, Subo and {Albrow}, Michael D. and {Chung}, Sun-Ju and {Han}, Cheongho and {Ryu}, Yoon-Hyun and {Shin}, In-Gu and {Shvartzvald}, Yossi and {Yang}, Hongjing and {Yee}, Jennifer C. and {Zang}, Weicheng and {Cha}, Sang-Mok and {Kim}, Dong-Jin and {Kim}, Seung-Lee and {Lee}, Chung-Uk and {Lee}, Dong-Joo and {Lee}, Yongseok and {Park}, Byeong-Gon and {Pogge}, Richard W.},
        title = "{Free-Floating Planets, the Einstein Desert, and 'OUMUAMUA}",
      journal = {Journal of Korean Astronomical Society},
         year = 2022,
        month = oct,
       volume = {55},
        pages = {173-194},
          doi = {10.5303/JKAS.2022.55.5.173},
archivePrefix = {arXiv},
       eprint = {2204.03269},
 primaryClass = {astro-ph.EP},
       adsurl = {https://ui.adsabs.harvard.edu/abs/2022JKAS...55..173G}
}

@ARTICLE{OB160263,
   author = {{Han}, C. and {Udalski}, A. and {Gould}, A. and {Bond}, I.~A. and 
	{and} and {Albrow}, M.~D. and {Chung}, S.-J. and {Jung}, Y.~K. and 
	{Ryu}, Y.-H. and {}, I.-G. and {Yee}, J.~C. and {Zhu}, W. and 
	{Cha}, S.-M. and {Kim}, S.-L. and {Kim}, D.-J. and {Lee}, C.-U. and 
	{Lee}, Y. and {Park}, B.-G. and {KMTNet Collaboration} and {Skowron}, J. and 
	{Mr{\'o}z}, P. and {Pietrukowicz}, P. and {Koz{\l}owski}, S. and 
	{Poleski}, R. and {Szyma{\'n}ski}, M.~K. and {Soszy{\'n}ski}, I. and 
	{Ulaczyk}, K. and {Pawlak}, M. and {OGLE Collaboration} and 
	{Abe}, F. and {Asakura}, Y. and {Barry}, R. and {Bennett}, D.~P. and 
	{Bhattacharya}, A. and {Donachie}, M. and {Evans}, P. and {Fukui}, A. and 
	{Hirao}, Y. and {Itow}, Y. and {Koshimoto}, N. and {Li}, M.~C.~A. and 
	{Ling}, C.~H. and {Masuda}, K. and {Matsubara}, Y. and {Muraki}, Y. and 
	{Nagakane}, M. and {Ohnishi}, K. and {Ranc}, C. and {Rattenbury}, N.~J. and 
	{Saito}, T. and {Sharan}, A. and {Sullivan}, D.~J. and {Sumi}, T. and 
	{Suzuki}, D. and {Tristram}, P.~J. and {Yamada}, T. and {Yamada}, T. and 
	{Yonehara}, A. and {The MOA Collaboration}},
    title = "{OGLE-2016-BLG-0263Lb: Microlensing Detection of a Very Low-mass Binary Companion through a Repeating Event Channel}",
  journal = {\aj},
archivePrefix = "arXiv",
   eprint = {1708.02727},
 primaryClass = "astro-ph.EP",
     year = 2017,
    month = oct,
   volume = 154,
      eid = {133},
    pages = {133},
      doi = {10.3847/1538-3881/aa859a},
   adsurl = {http://adsabs.harvard.edu/abs/2017AJ....154..133H}
}

@ARTICLE{Bensby2013,
   author = {{Bensby}, T. and {Yee}, J.~C. and {Feltzing}, S. and {Johnson}, J.~A. and 
	{Gould}, A. and {Cohen}, J.~G. and {Asplund}, M. and {Mel{\'e}ndez}, J. and 
	{Lucatello}, S. and {Han}, C. and {Thompson}, I. and {Gal-Yam}, A. and 
	{Udalski}, A. and {Bennett}, D.~P. and {Bond}, I.~A. and {Kohei}, W. and 
	{Sumi}, T. and {Suzuki}, D. and {Suzuki}, K. and {Takino}, S. and 
	{Tristram}, P. and {Yamai}, N. and {Yonehara}, A.},
    title = "{Chemical evolution of the Galactic bulge as traced by microlensed dwarf and subgiant stars. V. Evidence for a wide age distribution and a complex MDF}",
  journal = {\aap},
archivePrefix = "arXiv",
   eprint = {1211.6848},
     year = 2013,
    month = jan,
   volume = 549,
      eid = {A147},
    pages = {A147},
      doi = {10.1051/0004-6361/201220678},
   adsurl = {http://adsabs.harvard.edu/abs/2013A%26A...549A.147B}
}

@ARTICLE{KMTeventfinder,
   author = {{Kim}, D.-J. and {Kim}, H.-W. and {Hwang}, K.-H. and {Albrow}, M.~D. and 
	{Chung}, S.-J. and {Gould}, A. and {Han}, C. and {Jung}, Y.~K. and 
	{Ryu}, Y.-H. and {}, I.-G. and {Yee}, J.~C. and {Zhu}, W. and 
	{Cha}, S.-M. and {Kim}, S.-L. and {Lee}, C.-U. and {Lee}, D.-J. and 
	{Lee}, Y. and {Park}, B.-G. and {Pogge}, R.~W. and {The KMTNet Collaboration}
	},
    title = "{Korea Microlensing Telescope Network Microlensing Events from 2015: Event-finding Algorithm, Vetting, and Photometry}",
  journal = {\aj},
archivePrefix = "arXiv",
   eprint = {1703.06883},
 primaryClass = "astro-ph.EP",
     year = 2018,
    month = feb,
   volume = 155,
      eid = {76},
    pages = {76},
      doi = {10.3847/1538-3881/aaa47b},
   adsurl = {http://adsabs.harvard.edu/abs/2018AJ....155...76K}
}

@ARTICLE{Gould1992,
   author = {{Gould}, A.},
    title = "{Extending the MACHO search to about 10 exp 6 solar masses}",
  journal = {\apj},
     year = 1992,
    month = jun,
   volume = 392,
    pages = {442-451},
      doi = {10.1086/171443},
   adsurl = {http://adsabs.harvard.edu/abs/1992ApJ...392..442G}
}

@ARTICLE{Griest1998,
   author = {{Griest}, K. and {Safizadeh}, N.},
    title = "{The Use of High-Magnification Microlensing Events in Discovering Extrasolar Planets}",
  journal = {\apj},
   eprint = {astro-ph/9710342},
     year = 1998,
    month = jun,
   volume = 500,
    pages = {37-50},
      doi = {10.1086/305729},
   adsurl = {http://adsabs.harvard.edu/abs/1998ApJ...500...37G}
}

@ARTICLE{Kroupa2001,
   author = {{Kroupa}, P.},
    title = "{On the variation of the initial mass function}",
  journal = {\mnras},
   eprint = {astro-ph/0009005},
     year = 2001,
    month = apr,
   volume = 322,
    pages = {231-246},
      doi = {10.1046/j.1365-8711.2001.04022.x},
   adsurl = {http://adsabs.harvard.edu/abs/2001MNRAS.322..231K}
}

@ARTICLE{Yang_TLC,
       author = {{Yang}, Hongjing and {Yee}, Jennifer C. and {Hwang}, Kyu-Ha and {Qian}, Qiyue and {Bond}, Ian A. and {Gould}, Andrew and {Hu}, Zhecheng and {Zhang}, Jiyuan and {Mao}, Shude and {Zhu}, Wei and {Albrow}, Michael D. and {Chung}, Sun-Ju and {Kim}, Seung-Lee and {Park}, Byeong-Gon and {Han}, Cheongho and {Jung}, Youn Kil and {Ryu}, Yoon-Hyun and {Shin}, In-Gu and {Shvartzvald}, Yossi and {Cha}, Sang-Mok and {Kim}, Dong-Jin and {Kim}, Hyoun-Woo and {Lee}, Chung-Uk and {Lee}, Dong-Joo and {Lee}, Yongseok and {Pogge}, Richard W. and {Zang}, Weicheng and {Abe}, Fumio and {Barry}, Richard and {Bennett}, David P. and {Bhattacharya}, Aparna and {Donachie}, Martin and {Fujii}, Hirosane and {Fukui}, Akihiko and {Hirao}, Yuki and {Itow}, Yoshitaka and {Kirikawa}, Rintaro and {Kondo}, Iona and {Koshimoto}, Naoki and {Silva}, Stela Ishitani and {Li}, Man Cheung Alex and {Matsubara}, Yutaka and {Muraki}, Yasushi and {Suzuki}, Daisuke and {Tristram}, Paul J. and {Yonehara}, Atsunori and {Ranc}, Cl{\'e}ment and {Miyazaki}, Shota and {Olmschenk}, Greg and {Rattenbury}, Nicholas J. and {Satoh}, Yuki and {Shoji}, Hikaru and {Sumi}, Takahiro and {Tanaka}, Yuzuru and {Yamawaki}, Tsubasa},
        title = "{Systematic reanalysis of KMTNet microlensing events, paper I: Updates of the photometry pipeline and a new planet candidate}",
      journal = {\mnras},
         year = 2024,
        month = feb,
       volume = {528},
       number = {1},
        pages = {11-27},
          doi = {10.1093/mnras/stad3672},
archivePrefix = {arXiv},
       eprint = {2311.04876},
 primaryClass = {astro-ph.EP},
       adsurl = {https://ui.adsabs.harvard.edu/abs/2024MNRAS.528...11Y}
}

@ARTICLE{Yang_TLC2,
       author = {{Yang}, Hongjing and {Yee}, Jennifer C. and {Zhang}, Jiyuan and {Lee}, Chung-Uk and {Kim}, Dong-Jin and {Bond}, Ian A. and {Udalski}, Andrzej and {Hwang}, Kyu-Ha and {Zang}, Weicheng and {Qian}, Qiyue and {Gould}, Andrew and {Mao}, Shude and {Albrow}, Michael D. and {Chung}, Sun-Ju and {Han}, Cheongho and {Jung}, Youn Kil and {Ryu}, Yoon-Hyun and {Shin}, In-Gu and {Shvartzvald}, Yossi and {Cha}, Sang-Mok and {Kim}, Hyoun-Woo and {Kim}, Seung-Lee and {Lee}, Dong-Joo and {Lee}, Yongseok and {Park}, Byeong-Gon and {Pogge}, Richard W. and {KMTNet Collaboration} and {Abe}, Fumio and {Bando}, Ken and {Bennett}, David P. and {Bhattacharya}, Aparna and {Fukui}, Akihiko and {Hamada}, Ryusei and {Hamada}, Shunya and {Hamasaki}, Naoto and {Hirao}, Yuki and {Ishitani Silva}, Stela and {Itow}, Yoshitaka and {Koshimoto}, Naoki and {Matsubara}, Yutaka and {Miyazaki}, Shota and {Muraki}, Yasushi and {Nagai}, Tutumi and {Nunota}, Kansuke and {Olmschenk}, Greg and {Ranc}, Cl{\'e}ment and {Rattenbury}, Nicholas J. and {Satoh}, Yuki and {Sumi}, Takahiro and {Suzuki}, Daisuke and {Terry}, Sean K. and {Tristram}, Paul. J. and {Vandorou}, Aikaterini and {Yama}, Hibiki and {MOA Collaboration} and {Mr{\'o}z}, Przemek and {Skowron}, Jan and {Poleski}, Radoslaw and {Szyma{\'n}ski}, Micha{\l} K. and {Soszy{\'n}ski}, Igor and {Pietrukowicz}, Pawe{\l} and {Koz{\l}owski}, Szymon and {Ulaczyk}, Krzysztof and {Rybicki}, Krzysztof A. and {Iwanek}, Patryk and {Wrona}, Marcin and {OGLE Collaboration}},
        title = "{Systematic Reanalysis of KMTNet Microlensing Events. II. Two New Planets in Giant-source Events}",
      journal = {\aj},
         year = 2025,
        month = jun,
       volume = {169},
       number = {6},
          eid = {295},
        pages = {295},
          doi = {10.3847/1538-3881/adc73e},
archivePrefix = {arXiv},
       eprint = {2503.19471},
 primaryClass = {astro-ph.EP},
       adsurl = {https://ui.adsabs.harvard.edu/abs/2025AJ....169..295Y}
}

@ARTICLE{HSTCMD,
   author = {{Holtzman}, J.~A. and {Watson}, A.~M. and {Baum}, W.~A. and 
	{Grillmair}, C.~J. and {Groth}, E.~J. and {Light}, R.~M. and 
	{Lynds}, R. and {O'Neil}, Jr., E.~J.},
    title = "{The Luminosity Function and Initial Mass Function in the Galactic Bulge}",
  journal = {\aj},
   eprint = {astro-ph/9801321},
     year = 1998,
    month = may,
   volume = 115,
    pages = {1946-1957},
      doi = {10.1086/300336},
   adsurl = {http://adsabs.harvard.edu/abs/1998AJ....115.1946H}
}

@ARTICLE{MB07192,
   author = {{Bennett}, D.~P. and {Bond}, I.~A. and {Udalski}, A. and {Sumi}, T. and 
	{Abe}, F. and {Fukui}, A. and {Furusawa}, K. and {Hearnshaw}, J.~B. and 
	{Holderness}, S. and {Itow}, Y. and {Kamiya}, K. and {Korpela}, A.~V. and 
	{Kilmartin}, P.~M. and {Lin}, W. and {Ling}, C.~H. and {Masuda}, K. and 
	{Matsubara}, Y. and {Miyake}, N. and {Muraki}, Y. and {Nagaya}, M. and 
	{Okumura}, T. and {Ohnishi}, K. and {Perrott}, Y.~C. and {Rattenbury}, N.~J. and 
	{Sako}, T. and {Saito}, T. and {Sato}, S. and {Skuljan}, L. and 
	{Sullivan}, D.~J. and {Sweatman}, W.~L. and {Tristram}, P.~J. and 
	{Yock}, P.~C.~M. and {Kubiak}, M. and {Szyma{\'n}ski}, M.~K. and 
	{Pietrzy{\'n}ski}, G. and {Soszy{\'n}ski}, I. and {Szewczyk}, O. and 
	{Wyrzykowski}, {\L}. and {Ulaczyk}, K. and {Batista}, V. and 
	{Beaulieu}, J.~P. and {Brillant}, S. and {Cassan}, A. and {Fouqu{\'e}}, P. and 
	{Kervella}, P. and {Kubas}, D. and {Marquette}, J.~B.},
    title = "{A Low-Mass Planet with a Possible Sub-Stellar-Mass Host in Microlensing Event MOA-2007-BLG-192}",
  journal = {\apj},
archivePrefix = "arXiv",
   eprint = {0806.0025},
     year = 2008,
    month = sep,
   volume = 684,
    pages = {663-683},
      doi = {10.1086/589940},
   adsurl = {http://adsabs.harvard.edu/abs/2008ApJ...684..663B}
}

@ARTICLE{MatthewWFIRSTI,
       author = {{Penny}, Matthew T. and {Gaudi}, B. Scott and {Kerins}, Eamonn and
         {Rattenbury}, Nicholas J. and {Mao}, Shude and {Robin}, Annie C. and
         {Calchi Novati}, Sebastiano},
        title = "{Predictions of the WFIRST Microlensing Survey. I. Bound Planet Detection Rates}",
      journal = {\apjs},
         year = "2019",
        month = "Mar",
       volume = {241},
       number = {1},
          eid = {3},
        pages = {3},
          doi = {10.3847/1538-4365/aafb69},
archivePrefix = {arXiv},
       eprint = {1808.02490},
 primaryClass = {astro-ph.EP},
       adsurl = {https://ui.adsabs.harvard.edu/abs/2019ApJS..241....3P}
}

@ARTICLE{KMTAF,
       author = {{Kim}, Hyoun-Woo and {Hwang}, Kyu-Ha and {Shvartzvald}, Yossi and
         {Yee}, Jennifer C. and {Albrow}, Michael D. and {Cha}, Sang-Mok and
         {Chung}, Sun-Ju and {Gould}, Andrew and {Han}, Cheongho and
         {Jung}, Youn Kil and {Kim}, Dong-Jin and {Kim}, Seung-Lee and
         {Lee}, Chung-Uk and {Lee}, Dong-Joo and {Lee}, Yongseok and
         {Park}, Byeong-Gon and {Pogge}, Richard W. and {Ryu}, Yoon-Hyun and
         {Shin}, In-Gu and {Zang}, Weicheng},
        title = "{The Korea Microlensing Telescope Network (KMTNet) Alert Algorithm and Alert System}",
      journal = {arXiv e-prints},
         year = 2018,
        month = jun,
          eid = {arXiv:1806.07545},
        pages = {arXiv:1806.07545},
archivePrefix = {arXiv},
       eprint = {1806.07545},
 primaryClass = {astro-ph.IM},
       adsurl = {https://ui.adsabs.harvard.edu/abs/2018arXiv180607545K}
}

@ARTICLE{MB11293,
       author = {{Yee}, J.~C. and {Shvartzvald}, Y. and {Gal-Yam}, A. and {Bond}, I.~A. and
         {Udalski}, A. and {Koz{\l}owski}, S. and {Han}, C. and {Gould}, A. and
         {Skowron}, J. and {Suzuki}, D. and {Abe}, F. and {Bennett}, D.~P. and
         {Botzler}, C.~S. and {Chote}, P. and {Freeman}, M. and {Fukui}, A. and
         {Furusawa}, K. and {Itow}, Y. and {Kobara}, S. and {Ling}, C.~H. and
         {Masuda}, K. and {Matsubara}, Y. and {Miyake}, N. and {Muraki}, Y. and
         {Ohmori}, K. and {Ohnishi}, K. and {Rattenbury}, N.~J. and
         {Saito}, To. and {Sullivan}, D.~J. and {Sumi}, T. and {Suzuki}, K. and
         {Sweatman}, W.~L. and {Takino}, S. and {Tristram}, P.~J. and
         {Wada}, K. and {MOA Collaboration} and {Szyma{\'n}ski}, M.~K. and
         {Kubiak}, M. and {Pietrzy{\'n}ski}, G. and {Soszy{\'n}ski}, I. and
         {Poleski}, R. and {Ulaczyk}, K. and {Wyrzykowski}, {\L}. and
         {Pietrukowicz}, P. and {OGLE Collaboration} and {Allen}, W. and
         {Almeida}, L.~A. and {Batista}, V. and {Bos}, M. and {Christie}, G. and
         {DePoy}, D.~L. and {Dong}, Subo and {Drummond}, J. and {Finkelman}, I. and
         {Gaudi}, B.~S. and {Gorbikov}, E. and {Henderson}, C. and
         {Higgins}, D. and {Jablonski}, F. and {Kaspi}, S. and {Manulis}, I. and
         {Maoz}, D. and {McCormick}, J. and {McGregor}, D. and
         {Monard}, L.~A.~G. and {Moorhouse}, D. and {Mu{\~n}oz}, J.~A. and
         {Natusch}, T. and {Ngan}, H. and {Ofek}, E. and {Pogge}, R.~W. and
         {Santallo}, R. and {Tan}, T. -G. and {Thornley}, G. and {Shin}, I. -G. and
         {Choi}, J. -Y. and {Park}, S. -Y. and {Lee}, C. -U. and {Koo}, J. -R. and
         {{\ensuremath{\mu}}FUN Collaboration}},
        title = "{MOA-2011-BLG-293Lb: A Test of Pure Survey Microlensing Planet Detections}",
      journal = {\apj},
         year = 2012,
        month = aug,
       volume = {755},
       number = {2},
          eid = {102},
        pages = {102},
          doi = {10.1088/0004-637X/755/2/102},
archivePrefix = {arXiv},
       eprint = {1201.1002},
 primaryClass = {astro-ph.EP},
       adsurl = {https://ui.adsabs.harvard.edu/abs/2012ApJ...755..102Y}
}

@ARTICLE{Tomaney1996,
       author = {{Tomaney}, Austin B. and {Crotts}, Arlin P.~S.},
        title = "{Expanding the Realm of Microlensing Surveys with Difference Image Photometry}",
      journal = {\aj},
         year = "1996",
        month = "Dec",
       volume = {112},
        pages = {2872},
          doi = {10.1086/118228},
archivePrefix = {arXiv},
       eprint = {astro-ph/9610066},
 primaryClass = {astro-ph},
       adsurl = {https://ui.adsabs.harvard.edu/abs/1996AJ....112.2872T}
}

@ARTICLE{OB120950,
       author = {{Bhattacharya}, A. and {Beaulieu}, J. -P. and {Bennett}, D.~P. and
         {Anderson}, J. and {Koshimoto}, N. and {Lu}, J.~R. and {Batista}, V. and
         {Blackman}, J.~W. and {Bond}, I.~A. and {Fukui}, A. and
         {Henderson}, C.~B. and {Hirao}, Y. and {Marquette}, J.~B. and
         {Mroz}, P. and {Ranc}, C. and {Udalski}, A.},
        title = "{WFIRST Exoplanet Mass-measurement Method Finds a Planetary Mass of 39 {\ensuremath{\pm}} 8 M $_{{\ensuremath{\oplus}}}$ for OGLE-2012-BLG-0950Lb}",
      journal = {\aj},
         year = "2018",
        month = "Dec",
       volume = {156},
       number = {6},
          eid = {289},
        pages = {289},
          doi = {10.3847/1538-3881/aaed46},
archivePrefix = {arXiv},
       eprint = {1809.02654},
 primaryClass = {astro-ph.EP},
       adsurl = {https://ui.adsabs.harvard.edu/abs/2018AJ....156..289B}
}

@ARTICLE{GouldOort,
       author = {{Gould}, Andrew},
        title = "{Microlensing by Kuiper, Oort, and Free-Floating Planets}",
      journal = {Journal of Korean Astronomical Society},
         year = 2016,
        month = aug,
       volume = {49},
       number = {4},
        pages = {123-126},
          doi = {10.5303/JKAS.2016.49.4.123},
archivePrefix = {arXiv},
       eprint = {1603.03773},
 primaryClass = {astro-ph.EP},
       adsurl = {https://ui.adsabs.harvard.edu/abs/2016JKAS...49..123G}
}

@ARTICLE{Hanrepeating,
       author = {{Han}, Cheongho},
        title = "{Expansion of Planet Detection Methods in Next-Generation Microlensing Surveys}",
      journal = {\apj},
         year = 2007,
        month = dec,
       volume = {670},
       number = {2},
        pages = {1361-1366},
          doi = {10.1086/522024},
archivePrefix = {arXiv},
       eprint = {0707.4211},
 primaryClass = {astro-ph},
       adsurl = {https://ui.adsabs.harvard.edu/abs/2007ApJ...670.1361H}
}

@ARTICLE{OB08092,
       author = {{Poleski}, Rados{\l}aw and {Skowron}, Jan and {Udalski}, Andrzej and
         {Han}, Cheongho and {Koz{\l}owski}, Szymon and
         {Wyrzykowski}, {\L}ukasz and {Dong}, Subo and
         {Szyma{\'n}ski}, Micha{\l} K. and {Kubiak}, Marcin and
         {Pietrzy{\'n}ski}, Grzegorz and {Soszy{\'n}ski}, Igor and
         {Ulaczyk}, Krzysztof and {Pietrukowicz}, Pawe{\l} and {Gould}, Andrew},
        title = "{Triple Microlens OGLE-2008-BLG-092L: Binary Stellar System with a Circumprimary Uranus-type Planet}",
      journal = {\apj},
         year = 2014,
        month = nov,
       volume = {795},
       number = {1},
          eid = {42},
        pages = {42},
          doi = {10.1088/0004-637X/795/1/42},
archivePrefix = {arXiv},
       eprint = {1408.6223},
 primaryClass = {astro-ph.EP},
       adsurl = {https://ui.adsabs.harvard.edu/abs/2014ApJ...795...42P}
}

@ARTICLE{OB190960,
       author = {{Yee}, Jennifer C. and {Zang}, Weicheng and {Udalski}, Andrzej and {Ryu}, Yoon-Hyun and {Green}, Jonathan and {Hennerley}, Steve and {Marmont}, Andrew and {Sumi}, Takahiro and {Mao}, Shude and {Gromadzki}, Mariusz and {Mr{\'o}z}, Przemek and {Skowron}, Jan and {Poleski}, Radoslaw and {Szyma{\'n}ski}, Micha{\l} K. and {Soszy{\'n}ski}, Igor and {Pietrukowicz}, Pawe{\l} and {Koz{\l}owski}, Szymon and {Ulaczyk}, Krzysztof and {Rybicki}, Krzysztof A. and {Iwanek}, Patryk and {Wrona}, Marcin and {Albrow}, Michael D. and {Chung}, Sun-Ju and {Gould}, Andrew and {Han}, Cheongho and {Hwang}, Kyu-Ha and {Jung}, Youn Kil and {Kim}, Hyoun-Woo and {Shin}, In-Gu and {Shvartzvald}, Yossi and {Cha}, Sang-Mok and {Kim}, Dong-Jin and {Kim}, Seung-Lee and {Lee}, Chung-Uk and {Lee}, Dong-Joo and {Lee}, Yongseok and {Park}, Byeong-Gon and {Pogge}, Richard W. and {Bachelet}, Etienne and {Christie}, Grant and {Hundertmark}, Markus P.~G. and {Maoz}, Dan and {McCormick}, Jennie and {Natusch}, Tim and {Penny}, Matthew T. and {Street}, Rachel A. and {Tsapras}, Yiannis and {Beichman}, Charles A. and {Bryden}, Geoffery and {Novati}, Sebastiano Calchi and {Carey}, Sean and {Gaudi}, B. Scott and {Henderson}, Calen B. and {Johnson}, Samson and {Zhu}, Wei and {Bond}, Ian A. and {Abe}, Fumio and {Barry}, Richard and {Bennett}, David P. and {Bhattacharya}, Aparna and {Donachie}, Martin and {Fujii}, Hirosane and {Fukui}, Akihiko and {Hirao}, Yuki and {Silva}, Stela Ishitani and {Itow}, Yoshitaka and {Kirikawa}, Rintaro and {Kondo}, Iona and {Koshimoto}, Naoki and {Alex Li}, Man Cheung and {Matsubara}, Yutaka and {Muraki}, Yasushi and {Miyazaki}, Shota and {Olmschenk}, Greg and {Ranc}, Cl{\'e}ment and {Rattenbury}, Nicholas J. and {Satoh}, Yuki and {Shoji}, Hikaru and {Suzuki}, Daisuke and {Tanaka}, Yuzuru and {Tristram}, Paul J. and {Yamawaki}, Tsubasa and {Yonehara}, Atsunori and {MOA Collaboration}},
        title = "{OGLE-2019-BLG-0960 Lb: the Smallest Microlensing Planet}",
      journal = {\aj},
         year = 2021,
        month = nov,
       volume = {162},
       number = {5},
          eid = {180},
        pages = {180},
          doi = {10.3847/1538-3881/ac1582},
archivePrefix = {arXiv},
       eprint = {2101.04696},
 primaryClass = {astro-ph.EP},
       adsurl = {https://ui.adsabs.harvard.edu/abs/2021AJ....162..180Y}
}

@ARTICLE{KB200414,
       author = {{Zang}, Weicheng and {Han}, Cheongho and {Kondo}, Iona and {Yee}, Jennifer C. and {Lee}, Chung-Uk and {Gould}, Andrew and {Mao}, Shude and {de Almeida}, Leandro and {Shvartzvald}, Yossi and {Zhang}, Xiangyu and {Albrow}, Michael D. and {Chung}, Sun-Ju and {Hwang}, Kyu-Ha and {Jung}, Youn Kil and {Ryu}, Yoon-Hyun and {Shin}, In-Gu and {Cha}, Sang-Mok and {Kim}, Dong-Jin and {Kim}, Hyoun-Woo and {Kim}, Seung-Lee and {Lee}, Dong-Joo and {Lee}, Yongseok and {Park}, Byeong-Gon and {Pogge}, Richard W. and {Drummond}, John and {Tan}, Thiam-Guan and {Nascimento J{\'u}nior}, Jos{\'e} Dias do and {Maoz}, Dan and {Penny}, Matthew T. and {Zhu}, Wei and {Bond}, Ian A. and {Abe}, Fumio and {Barry}, Richard and {Bennett}, David P. and {Bhattacharya}, Aparna and {Donachie}, Martin and {Fujii}, Hirosane and {Fukui}, Akihiko and {Hirao}, Yuki and {Itow}, Yoshitaka and {Kirikawa}, Rintaro and {Koshimoto}, Naoki and {Alex Li}, Man Cheung and {Matsubara}, Yutaka and {Muraki}, Yasushi and {Miyazaki}, Shota and {Olmschenk}, Greg and {Ranc}, Cl{\'e}ment and {Rattenbury}, Nicholas J. and {Satoh}, Yuki and {Shoji}, Hikaru and {Silva}, Stela Ishitani and {Sumi}, Takahiro and {Suzuki}, Daisuke and {Tanaka}, Yuzuru and {Tristram}, Paul J. and {Yamawaki}, Tsubasa and {Yonehara}, Atsunori and {Petric}, Andreea and {Burdullis}, Todd and {Fouqu{\'e}}, Pascal},
        title = "{An Earth-mass planet in a time of COVID-19: KMT-2020-BLG-0414Lb}",
      journal = {Research in Astronomy and Astrophysics},
         year = 2021,
        month = nov,
       volume = {21},
       number = {9},
          eid = {239},
        pages = {239},
          doi = {10.1088/1674-4527/21/9/239},
archivePrefix = {arXiv},
       eprint = {2103.01896},
 primaryClass = {astro-ph.EP},
       adsurl = {https://ui.adsabs.harvard.edu/abs/2021RAA....21..239Z}
}

@ARTICLE{emcee2,
       author = {{Goodman}, Jonathan and {Weare}, Jonathan},
        title = "{Ensemble samplers with affine invariance}",
      journal = {Communications in Applied Mathematics and Computational Science},
         year = 2010,
        month = jan,
       volume = {5},
       number = {1},
        pages = {65-80},
          doi = {10.2140/camcos.2010.5.65},
       adsurl = {https://ui.adsabs.harvard.edu/abs/2010CAMCS...5...65G}
}

@ARTICLE{OGLE_wide,
       author = {{Poleski}, R. and {Skowron}, J. and {Mr{\'o}z}, P. and {Udalski}, A. and {Szyma{\'n}ski}, M.~K. and {Pietrukowicz}, P. and {Ulaczyk}, K. and {Rybicki}, K. and {Iwanek}, P. and {Wrona}, M. and {Gromadzki}, M.},
        title = "{Wide-Orbit Exoplanets are Common. Analysis of Nearly 20 Years of OGLE Microlensing Survey Data}",
      journal = {\actaa},
         year = 2021,
        month = mar,
       volume = {71},
       number = {1},
        pages = {1-23},
          doi = {10.32023/0001-5237/71.1.1},
archivePrefix = {arXiv},
       eprint = {2104.02079},
 primaryClass = {astro-ph.EP},
       adsurl = {https://ui.adsabs.harvard.edu/abs/2021AcA....71....1P}
}

@ARTICLE{CMST,
       author = {{Gould}, Andrew and {Zang}, Wei-Cheng and {Mao}, Shude and {Dong}, Su-Bo},
        title = "{Masses for free-floating planets and dwarf planets}",
      journal = {Research in Astronomy and Astrophysics},
         year = 2021,
        month = aug,
       volume = {21},
       number = {6},
          eid = {133},
        pages = {133},
          doi = {10.1088/1674-4527/21/6/133},
archivePrefix = {arXiv},
       eprint = {2010.09671},
 primaryClass = {astro-ph.EP},
       adsurl = {https://ui.adsabs.harvard.edu/abs/2021RAA....21..133G}
}

@ARTICLE{ET,
       author = {{Ge}, Jian and {Zhang}, Hui and {Zang}, Weicheng and {Deng}, Hongping and {Mao}, Shude and {Xie}, Ji-Wei and {Liu}, Hui-Gen and {Zhou}, Ji-Lin and {Willis}, Kevin and {Huang}, Chelsea and {Howell}, Steve B. and {Feng}, Fabo and {Zhu}, Jiapeng and {Yao}, Xinyu and {Liu}, Beibei and {Aizawa}, Masataka and {Zhu}, Wei and {Li}, Ya-Ping and {Ma}, Bo and {Ye}, Quanzhi and {Yu}, Jie and {Xiang}, Maosheng and {Yu}, Cong and {Liu}, Shangfei and {Yang}, Ming and {Wang}, Mu-Tian and {Shi}, Xian and {Fang}, Tong and {Zong}, Weikai and {Liu}, Jinzhong and {Zhang}, Yu and {Zhang}, Liyun and {El-Badry}, Kareem and {Shen}, Rongfeng and {Tam}, Pak-Hin Thomas and {Hu}, Zhecheng and {Yang}, Yanlv and {Zou}, Yuan-Chuan and {Wu}, Jia-Li and {Lei}, Wei-Hua and {Wei}, Jun-Jie and {Wu}, Xue-Feng and {Sun}, Tian-Rui and {Wang}, Fa-Yin and {Zhang}, Bin-Bin and {Xu}, Dong and {Yang}, Yuan-Pei and {Li}, Wen-Xiong and {Xiang}, Dan-Feng and {Wang}, Xiaofeng and {Wang}, Tinggui and {Zhang}, Bing and {Jia}, Peng and {Yuan}, Haibo and {Zhang}, Jinghua and {Xuesong Wang}, Sharon and {Gan}, Tianjun and {Wang}, Wei and {Zhao}, Yinan and {Liu}, Yujuan and {Wei}, Chuanxin and {Kang}, Yanwu and {Yang}, Baoyu and {Qi}, Chao and {Liu}, Xiaohua and {Zhang}, Quan and {Zhu}, Yuji and {Zhou}, Dan and {Zhang}, Congcong and {Yu}, Yong and {Zhang}, Yongshuai and {Li}, Yan and {Tang}, Zhenghong and {Wang}, Chaoyan and {Wang}, Fengtao and {Li}, Wei and {Cheng}, Pengfei and {Shen}, Chao and {Li}, Baopeng and {Pan}, Yue and {Yang}, Sen and {Gao}, Wei and {Song}, Zongxi and {Wang}, Jian and {Zhang}, Hongfei and {Chen}, Cheng and {Wang}, Hui and {Zhang}, Jun and {Wang}, Zhiyue and {Zeng}, Feng and {Zheng}, Zhenhao and {Zhu}, Jie and {Guo}, Yingfan and {Zhang}, Yihao and {Li}, Yudong and {Wen}, Lin and {Feng}, Jie and {Chen}, Wen and {Chen}, Kun and {Han}, Xingbo and {Yang}, Yingquan and {Wang}, Haoyu and {Duan}, Xuliang and {Huang}, Jiangjiang and {Liang}, Hong and {Bi}, Shaolan and {Gai}, Ning and {Ge}, Zhishuai and {Guo}, Zhao and {Huang}, Yang and {Li}, Gang and {Li}, Haining and {Li}, Tanda and {Yuxi} and {Lu} and {Rix}, Hans-Walter and {Shi}, Jianrong and {Song}, Fen and {Tang}, Yanke and {Ting}, Yuan-Sen and {Wu}, Tao and {Wu}, Yaqian and {Yang}, Taozhi and {Yin}, Qing-Zhu and {Gould}, Andrew and {Lee}, Chung-Uk and {Dong}, Subo and {Yee}, Jennifer C. and {Shvartzvald}, Yossi and {Yang}, Hongjing and {Kuang}, Renkun and {Zhang}, Jiyuan and {Liao}, Shilong and {Qi}, Zhaoxiang and {Yang}, Jun and {Zhang}, Ruisheng and {Jiang}, Chen and {Ou}, Jian-Wen and {Li}, Yaguang and {Beck}, Paul and {Bedding}, Timothy R. and {Campante}, Tiago L. and {Chaplin}, William J. and {Christensen-Dalsgaard}, J{\o}rgen and {Garc{\'\i}a}, Rafael A. and {Gaulme}, Patrick and {Gizon}, Laurent and {Hekker}, Saskia and {Huber}, Daniel and {Khanna}, Shourya and {Li}, Yan and {Mathur}, Savita and {Miglio}, Andrea and {Mosser}, Beno{\^\i}t and {Ong}, J.~M. Joel and {Santos}, {\^A}ngela R.~G. and {Stello}, Dennis and {Bowman}, Dominic M. and {Lares-Martiz}, Mariel and {Murphy}, Simon and {Niu}, Jia-Shu and {Ma}, Xiao-Yu and {Moln{\'a}r}, L{\'a}szl{\'o} and {Fu}, Jian-Ning and {De Cat}, Peter and {Su}, Jie and {consortium}, the ET},
        title = "{ET White Paper: To Find the First Earth 2.0}",
      journal = {arXiv e-prints},
         year = 2022,
        month = jun,
          eid = {arXiv:2206.06693},
        pages = {arXiv:2206.06693},
archivePrefix = {arXiv},
       eprint = {2206.06693},
 primaryClass = {astro-ph.IM},
       adsurl = {https://ui.adsabs.harvard.edu/abs/2022arXiv220606693G}
}

@ARTICLE{KB240792,
       author = {{Dong}, Subo and {Wu}, Zexuan and {Ryu}, Yoon-Hyun and {Udalski}, Andrzej and {Mr{\'o}z}, Przemek and {Rybicki}, Krzysztof A. and {Hodgkin}, Simon T. and {Wyrzykowski}, {\L}ukasz and {Eyer}, Laurent and {Bensby}, Thomas and {Chen}, Ping and {Wang}, Sharon X. and {Gould}, Andrew and {Yang}, Hongjing and {Albrow}, Michael D. and {Chung}, Sun-Ju and {Han}, Cheongho and {Hwang}, Kyu-Ha and {Jung}, Youn Kil and {Shin}, In-Gu and {Shvartzvald}, Yossi and {Yee}, Jennifer C. and {Zang}, Weicheng and {Kim}, Dong-Jin and {Lee}, Chung-Uk and {Park}, Byeong-Gon and {Poleski}, Rados{\l}aw and {Skowron}, Jan and {Szyma{\'n}ski}, Micha{\l} K. and {Soszy{\'n}ski}, Igor and {Pietrukowicz}, Pawe{\l} and {Koz{\l}owski}, Szymon and {Skowron}, Dorota M. and {Ulaczyk}, Krzysztof and {Gromadzki}, Mariusz and {Ratajczak}, Milena and {Iwanek}, Patryk and {Wrona}, Marcin and {Mr{\'o}z}, Mateusz J. and {Rixon}, Guy and {Harrison}, Diana L. and {Breedt}, Elm{\'e}},
        title = "{A free-floating-planet microlensing event caused by a Saturn-mass object}",
      journal = {Science},
         year = 2026,
        month = jan,
       volume = {391},
       number = {6780},
        pages = {96-99},
          doi = {10.1126/science.adv9266},
archivePrefix = {arXiv},
       eprint = {2601.00057},
 primaryClass = {astro-ph.EP},
       adsurl = {https://ui.adsabs.harvard.edu/abs/2026Sci...391...96D}
}

@ARTICLE{PARSEC,
       author = {{Bressan}, Alessandro and {Marigo}, Paola and {Girardi}, L{\'e}o. and {Salasnich}, Bernardo and {Dal Cero}, Claudia and {Rubele}, Stefano and {Nanni}, Ambra},
        title = "{PARSEC: stellar tracks and isochrones with the PAdova and TRieste Stellar Evolution Code}",
      journal = {\mnras},
         year = 2012,
        month = nov,
       volume = {427},
       number = {1},
        pages = {127-145},
          doi = {10.1111/j.1365-2966.2012.21948.x},
archivePrefix = {arXiv},
       eprint = {1208.4498},
 primaryClass = {astro-ph.SR},
       adsurl = {https://ui.adsabs.harvard.edu/abs/2012MNRAS.427..127B}
}

@ARTICLE{scipy,
       author = {{Virtanen}, Pauli and {Gommers}, Ralf and {Oliphant}, Travis E. and {Haberland}, Matt and {Reddy}, Tyler and {Cournapeau}, David and {Burovski}, Evgeni and {Peterson}, Pearu and {Weckesser}, Warren and {Bright}, Jonathan and {van der Walt}, St{\'e}fan J. and {Brett}, Matthew and {Wilson}, Joshua and {Millman}, K. Jarrod and {Mayorov}, Nikolay and {Nelson}, Andrew R.~J. and {Jones}, Eric and {Kern}, Robert and {Larson}, Eric and {Carey}, C.~J. and {Polat}, {\.I}lhan and {Feng}, Yu and {Moore}, Eric W. and {VanderPlas}, Jake and {Laxalde}, Denis and {Perktold}, Josef and {Cimrman}, Robert and {Henriksen}, Ian and {Quintero}, E.~A. and {Harris}, Charles R. and {Archibald}, Anne M. and {Ribeiro}, Ant{\^o}nio H. and {Pedregosa}, Fabian and {van Mulbregt}, Paul and {SciPy 1. 0 Contributors}},
        title = "{SciPy 1.0: fundamental algorithms for scientific computing in Python}",
      journal = {Nature Methods},
         year = 2020,
        month = feb,
       volume = {17},
        pages = {261-272},
          doi = {10.1038/s41592-019-0686-2},
archivePrefix = {arXiv},
       eprint = {1907.10121},
 primaryClass = {cs.MS},
       adsurl = {https://ui.adsabs.harvard.edu/abs/2020NatMe..17..261V}
}

@ARTICLE{Di1996_repeatevent,
       author = {{Di Stefano}, R. and {Mao}, S.},
        title = "{Do Microlensing Events Repeat?}",
      journal = {\apj},
         year = 1996,
        month = jan,
       volume = {457},
        pages = {93},
          doi = {10.1086/176713},
       adsurl = {https://ui.adsabs.harvard.edu/abs/1996ApJ...457...93D}
}

\end{CJK*}
\end{document}